\documentclass[useAMS,usenatbib]{mn2e}
\usepackage{natbib}
\usepackage{amsmath,amsfonts}
\usepackage{epsfig}
\usepackage{multirow}
\usepackage{mathptmx}
\usepackage{longtable}
\usepackage{graphicx}
\usepackage{subfig}
\usepackage{float}
\usepackage{lscape}
\usepackage{rotating}
\usepackage{pdflscape}
\usepackage{caption}

\def\reff@jnl#1{{\rm#1\/}}

\def\aj{\reff@jnl{AJ}}                  
\def\araa{\reff@jnl{ARA\&A}}            
\def\apj{\reff@jnl{ApJ}}                        
\def\apjl{\reff@jnl{ApJ}}               
\def\apjs{\reff@jnl{ApJS}}              
\def\ao{\reff@jnl{Appl.Optics}}         
\def\apss{\reff@jnl{Ap\&SS}}            
\def\aap{\reff@jnl{A\&A}}                       
\def\apjl{\reff@jnl{ApJ}}               
\def\aapr{\reff@jnl{A\&A~Rev.}}         
\def\aaps{\reff@jnl{A\&AS}}             
\def\azh{\reff@jnl{AZh}}                        
\def\baas{\reff@jnl{BAAS}}              
\def\jrasc{\reff@jnl{JRASC}}            
\def\memras{\reff@jnl{MmRAS}}           
\def\mnras{\reff@jnl{MNRAS}}            
\def\pra{\reff@jnl{Phys. Rev. A}}         
\def\prb{\reff@jnl{Phys. Rev. B}}         
\def\prc{\reff@jnl{Phys. Rev. C}}         
\def\prd{\reff@jnl{Phys. Rev. D}}         
\def\prl{\reff@jnl{Phys. Rev. Lett}}      
\def\pasp{\reff@jnl{PASP}}              
\def\pasj{\reff@jnl{PASJ}}              
\def\qjras{\reff@jnl{QJRAS}}            
\def\skytel{\reff@jnl{S\&T}}            
\def\solphys{\reff@jnl{Solar~Phys.}}    
\def\sovast{\reff@jnl{Soviet~Ast.}}     
\def\ssr{\reff@jnl{Space~Sci.Rev.}}     
\def\zap{\reff@jnl{ZAp}}                        
\def\nat{\reff@jnl{Nature}}             

\def\p#1by#2{{\partial{#1} \over \partial{#2}}}
\def\pp#1by#2#3{{\partial^2{#1} \over \partial{#2}\partial{#3}}}
\def\d#1by#2{{{\rm d}{#1} \over {\rm d}{#2}}}
\def\dd#1by#2#3{{{\rm d}^2{#1} \over {\rm d}{#2}{\rm d}{#3}}}

\makeatletter

\newcommand{\Rmnum}[1]{\expandafter\@slowromancap\romannumeral #1@}

\newcommand{\hi}{{\sc H\,i\,}}

\newcommand{\mic}{\,$\mu$m\,}
\newcommand{\kms}{\,km s$^{-1}$}
\makeatother
\begin{document}
\title[Coma cluster galaxies]{H-ATLAS: The Far-Infrared properties of galaxies in and around the Coma Cluster\thanks{{\it Herschel} is an ESA space observatory with science instruments provided by European-led Principal Investigator consortia and with important participation from NASA.}}
\author[Fuller et al.]{
C. Fuller$^{1}$,
J.I. Davies$^{1}$,
M.W.L. Smith$^{1}$,
E. Valiante$^{1}$,
S. Eales$^{1}$,
N. Bourne${^2}$,
\newauthor
L. Dunne$^{1,2}$, 
S. Dye${^4}$, 
C. Furlanetto${^3}$
E. Ibar$^{3}$,
R. Ivison$^{2,5}$ 
S. Maddox$^{1,2}$
\newauthor
A. Sansom$^{6}$,
M.~J.~Micha{\l}owski$^{2}$ and
T. Davis$^{1}$ \\
$^{1}$ School of Physics and Astronomy, Cardiff University, The Parade, Cardiff CF24 3AA, UK. email:Jonathan.Davies@astro.cf.ac.uk.\\
$^{2}$ Institute for Astronomy, The University of Edinburgh, Royal Observatory, Blackford Hill,
Edinburgh, EH9 3HJ, UK.\\
$^{3}$ Instituto de F\'isica y Astronom\'ia, Universidad de Valpara\'iso, Avda. Gran Breta\~na 1111,
Valpara\'iso, Chile.\\
$^{4}$ School of Physics and Astronomy, University of Nottingham, University Park, Nottingham
NG7 2RD, UK.\\
$^{5}$ European Southern Observatory, Karl-Schwarzschild-Strasse 2, 85748, Garching, Germany.\\
$^{6}$ University of Central Lancashire, Physics Dept., Corporation St., Preston, Lancs, PR1 6AD, UK. \\
}
\date{\today}
\maketitle

\begin{abstract}
We describe a far infrared survey of the Coma cluster and the galaxy filament it resides within.  Our survey covers an area of $\sim$150 deg$^2$ observed by $Herschel$ H-ATLAS in five bands  at 100, 160, 250, 350 and 500\,$\mu$m. The SDSS spectroscopic survey ($m_{r} \le 17.8)$ is used to define an area (within the Virial radius) and redshift selected ($4268 < v < 9700$ km s$^{-1}$) sample of 744 Coma cluster galaxies - the Coma Cluster Catalogue (CCC). For comparison we also define a sample of 951 galaxies in the connecting filament - the Coma Filament Catalogue (CFC).
The optical positions and parameters are used to define appropriate apertures to measure each galaxy's far infrared emission. We have detected 99 of 744 (13\,\%)  and 422 of 951 (44\,\%) of the cluster and filament galaxies in the SPIRE 250\,$\mu$m band.  We consider the relative detection rates of galaxies of different morphological types finding that it is only the S0/Sa population that shows clear differences between the cluster and filament. We find no differences between the dust masses and temperatures of cluster and filament galaxies with the exception of early type galaxy dust temperatures, which are significantly hotter in the cluster than in the filament (X-ray heating?). From a chemical evolution model we find no evidence for different evolutionary processes (gas loss or infall) between galaxies in the cluster and filament. \end{abstract}

\section{Introduction}

The Coma cluster is a local example of a giant relaxed structure. Coma is located at a distance of $\sim$\,100\,Mpc, has a mass of $\sim10^{15}$ M$_{\odot}$ and a virial radius $\sim$3 Mpc (\cite{girardi1998}, \cite{boselli06}). The cluster resides in a large filamentary structure known as the `Great Wall', which is an over density (galactic filament) of galaxies linking Coma with other groups and clusters including another nearby cluster A1367 \citep{ramella1992}.  The cluster resides at a high galactic latitude of $b$=88.0$^{o}$, making it ideal for extra-galactic studies at all wavelengths. Notably observed using the 48 inch reflector on Mount Wilson by ~\cite{zwicky51} it was not until 1983 that ~\cite{gmp} created the most cited optical catalogue of `Coma' galaxies. They catalogued over 6000 galaxies in the region of the cluster, but not all of these are cluster members as the origin of their data was photographic plate imaging, with no spectroscopy to distinguish cluster members from the foreground/background. For this reason an important first part of our project has been to produce an optical catalogue of Coma cluster members that can be used in the same way as that used for other nearby clusters such as Virgo (Virgo Cluster Catalogue (VCC), ~\cite{vcc}) and Fornax (Fornax Cluster Catalogue (FCC), ~\cite{ferguson90}).

Compared to a cluster like Virgo, Coma has a much more centrally concentrated distribution of galaxies and mass. The latter can be inferred from the X-ray emitting gas~\citep{colless95}, which is thought to be material originally lost by member galaxies, that now traces the gravitational potential of the cluster. The X-ray gas shows a mostly smooth, centrally concentrated distribution of mass. This X-ray emitting gas can also have a significant effect on galaxies as they plunge through the cluster at relatively high velocities ($\sim$1000 km s$^{-1}$). Ram pressure stripping~\citep{gunn72} and thermal evaporation~\citep{cowie77} act to remove atomic hydrogen from galaxies as bright as $L_{*}$ on time scales comparable to a cluster crossing time, which for Coma is at most a few Gyr ~\citep{boselli06}. Thus we might expect the star formation history of Coma cluster galaxies to be very different to those that reside outside or on the outskirts of the cluster (\cite{kennicutt1984}, \cite{donas1995}, \cite{gavazzi1998}, \cite{gavazzi1998}, \cite{gavazzi1998}, \cite{gavazzi2010}, \cite{cybulski2014}). Indeed, we find strong morphological segregation in Coma, with a large proportion of its members being early-type galaxies (see below). The inter-stellar medium (ISM) of the cluster late-type galaxies is observed to be affected by the cluster environment (\cite{giov1985}, \cite{gavazzi1989}, \cite{solanes2001}, \cite{casoli1996},  \cite{fumagalli2009}, \cite{boselli2014}, \cite{cortese10}, \cite{cortese12}), something that is supported by simulation (\cite{tonnesen07}, \cite{roediger08}). ~\citet{gavazzi06b} found that Coma galaxies are H\Rmnum{1}-deficient to $\sim$1.5 times the clusters Virial radius, and most  H\Rmnum{1} deficient at the X-ray centre of the cluster. Given this obviously environmental dependence on interstellar medium properties we will also study here for comparison, along with the Coma cluster galaxies, a sample of galaxies residing in the Great Wall,  well beyond the Virial radius (defined below) of the cluster.

Although this paper makes use of our optically selected catalogue of Coma (cluster) and Great Wall (filament) galaxies it is primarily concerned with Herschel Space Telescope observations at far-infrared wavelengths (see also, \cite{bicay1987}, \cite{contursi01}, \cite{bai2006}). These far-infrared observations are predominately measuring emission from relatively cold (10-50 K) cosmic dust. This cosmic dust has been shown to be a good tracer of star formation, molecular material and of metal content and hence galactic chemical evolution, which are themes we explore in this paper.

Due to extinction in the Earth's atmosphere far-infrared observations are limited to space based telescopes. Previously the first infrared space mission, IRAS~\citep{IRAS} detected 41 sources in Coma as part of its all sky survey (wavelengths 25-100 $\mu$m). The sample was almost entirely composed of late type galaxies in the periphery of the cluster. Subsequently using ISO~\citep{ISO}, which extended the far infrared wavelength coverage to $\sim$200\,$\mu$m),~\citet{contursi01} detected 11 Coma galaxies. They found that even though the galaxies where interacting with the intergalactic  medium, their dust properties seemed to be strangely un-affected by this. ISO's extended wavelength coverage also revealed a previously unseen cold dust component ($\sim$10K), but it was rather poorly constrained because of lack of data beyond 200$\mu$m.  The Spitzer Space Telescope~\citep{spitzer} did not extend the wavelength coverage of ISO, but it did improve sensitivity. \citet{edwards11} used Spitzer to observe multiple fields in Coma covering a total area of $\sim$3 $\deg^{2}$, approximately $1/9$ of the area traced by the Virial radius. They confirmed that star formation was strongly suppressed in Coma's core and observed some starburst galaxies in the south west, which they identified as a region where galaxies were currently falling into the cluster for the first time. 

The above instruments primarily allowed the study of what we would now refer to as warm dust (T$\approx$30 K). They were limited in their ability to constrain the amount of cold dust (T$<$30 K) as they did not provide accurate measurements of the Raleigh-Jeans tail of the modified blackbody emission, which typically has been shown to peak at $\sim$160\,$\mu$m~\citep{davies12a}.  

The $Herschel$ $Space$ $Telescope$ \citep{pilbratt10} and its instruments were designed to overcome this problem, enabling observations out to about 500 $\mu$m and hence a measurement of the total dust content of galaxies. These observations can be used to investigate the issues briefly outlined above. 

Previously ~\citet{Hickinbottom14} have published deep $Herschel$ observations of 1.75 sq deg of the Coma cluster core at 70, 100 and 160$\mu$m. We can now extend these observations both spatially and in wavelength range.

The Coma cluster was also observed as part of The $Herschel$ $Astrophysical$ $Terahertz$ $Large$ $Area$ $Survey$ (H-ATLAS)~\citep{eales10} and it is this data we use in this paper. This is the largest $Herschel$ FIR extra-galactic survey covering 570 deg$^{2}$ in 5 bands centred at 100, 160, 250, 350 and 500\,$\mu$m. With high Galactic latitudes the H-ATLAS fields are all located in areas of low galactic cirrus, Coma being part of the Northern Galactic Cap (NGP) field. Along with Coma there is also a wide spatial coverage of the surrounding areas well beyond the cluster's Virial radius into its connecting galactic filament. 

\section{The Optical selection of cluster and filament galaxies}
Unlike our previous work on the Virgo~\citep{Davies14} and Fornax~\citep{Fuller14} clusters there is no equivalent optical catalogue of Coma cluster galaxies\footnote{Similar data to that used here for the combined Coma and A1365 clusters along with galaxies in the "Great Wall" has previously been described in \cite{gavazzi2010}.}. Our previous papers on Virgo and Fornax used optical data from the VCC and FCC to distinguish cluster members from background galaxies. The often cited Coma catalogue of ~\citet{gmp} is actually just a list of all objects detected on a photographic plate with no additional means of separating cluster from non-cluster galaxies. Specific redshift surveys do exist for various parts of the Coma region,~\citep{kent82, colless96,geller99, castander01, mobasher01}, but the lack of homogeneity of these surveys precluded the generation of a unified redshift selected Coma cluster catalogue suitable for our purposes.

\begin{figure}
\centering
\includegraphics[width=\linewidth]{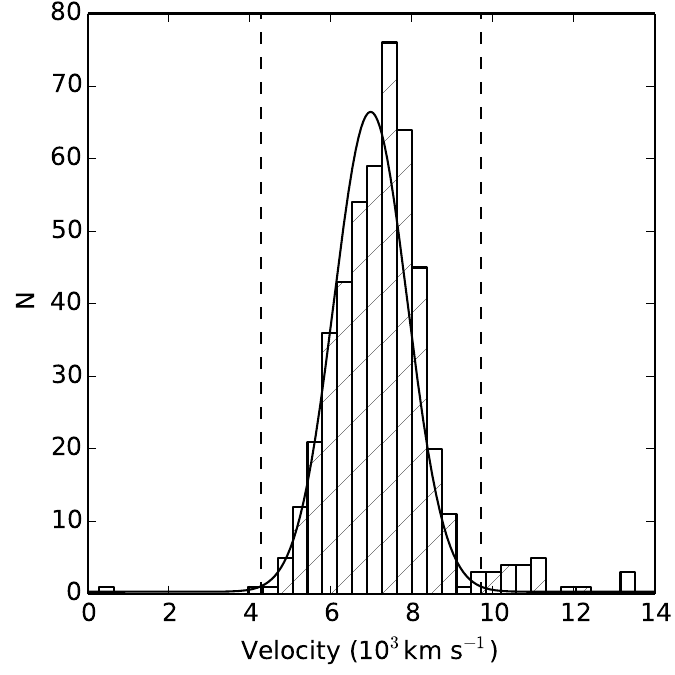}
\caption[Velocity distribution of the CCC galaxies]{The velocity distribution of the CCC galaxies, showing a clearly relaxed system. The vertical dashed lines represent the 3$\sigma$ velocity dispersion of the fitted Gaussian function (solid line).}
\label{fig:scematic}
\end{figure}

However, this region of the sky is now covered by both the spectroscopic and photometric SDSS surveys, and thus allows us to select a clearly defined optical catalogue with secure redshifts. This also enables us to select and isolate both Coma cluster and filament (`Great Wall') galaxies. Our hope is that this Coma Cluster Catalogue (CCC) will be as useful a tool for studies of nearby galaxies and clusters as both the VCC and FCC have already been.

The SDSS Main Galaxy Sample (MGS) is selected from the spectroscopic survey according to criteria discussed extensively in~\citet{strauss02}. Briefly, objects are selected if they are detected at 5$\sigma$ or above in the $r$ band. Galaxies are then separated from stars by deciding if they are point or extended sources and also stipulating that they have not been flagged as SATURATED, BRIGHT, or BLENDED. Finally the selected sources must have an $r$ band apparent magnitude brighter than  $m_{r}=17.8$\footnote{To confirm that the SDSS magnitudes we have used are consistent across the magnitude range we have checked them against the V band data available through $Goldmine$. We find a good one-to-one linear relationship across the whole range of magnitudes used.}.  
The SDSS spectrograph also had the physical limitation that two fibres could not be closer together than 55"  or $\sim$28 kpc at a Coma distance of 105 Mpc~\citep{boselli06}, but this was not a problem for the bright Coma cluster galaxies considered here ($m_{r}=17.8$ corresponds to an absolute magnitude of $\approx-17.3$ at the above Coma distance).
Our sample is selected from the MGS (SDSS DR10) by isolating the cluster and filament in both spatial and velocity extent.

Our initial selection was for all galaxies within a velocity range 3000 to 11000 km s$^{-1}$, with a spatial selection over the area of the Herschel NGP field. The velocity range is roughly symmetrical about the current value for Coma's mean velocity as listed in the Nasa Extragalactic Database\footnote{http://ned.ipac.caltech.edu} (NED) of $\sim 7000$\,km s$^{-1}$.

\subsection{Defining the cluster}\label{sec:CCC}
Many clusters are far from Virial equilibrium and so R$_{200}$, the radius at which the mass density drops to 1/200 of the critical density, is often used as a measure of their size. Coma however appears to be a dynamically mature cluster, making the Virial radius a less arbitrary choice in this particular case. A cluster in Virial equilibrium would have a velocity distribution perfectly traced by a Gaussian function. Figure~\ref{fig:scematic} (details below) shows that this is a good interpretation of the Coma data. 

We have defined the centre of Coma as the position of peak X-ray emission~\citep{colless95}. This X-ray emitting gas is a much more reliable indicator of the mass distribution of the cluster than the optical surface density of galaxies, as it better traces the total mass distribution.  Using this definition we measure the centre to be at: RA(J2000)=12h59m48.7s and DEC(J2000)=+27d58m50.0s. From the literature we have found four different derived values for the Virial radius of Coma~\citep{kubo07,geller99,hughes89,the86}. After adjusting to a common distance scale (H$_{0}=70$ km s$^{-1}$ Mpc$^{-1}$) we have used the mean of these four values for the Virial radius i.e. $3.1 \pm 0.7$ Mpc. 

\begin{figure*}
\centering
\includegraphics[width=\linewidth]{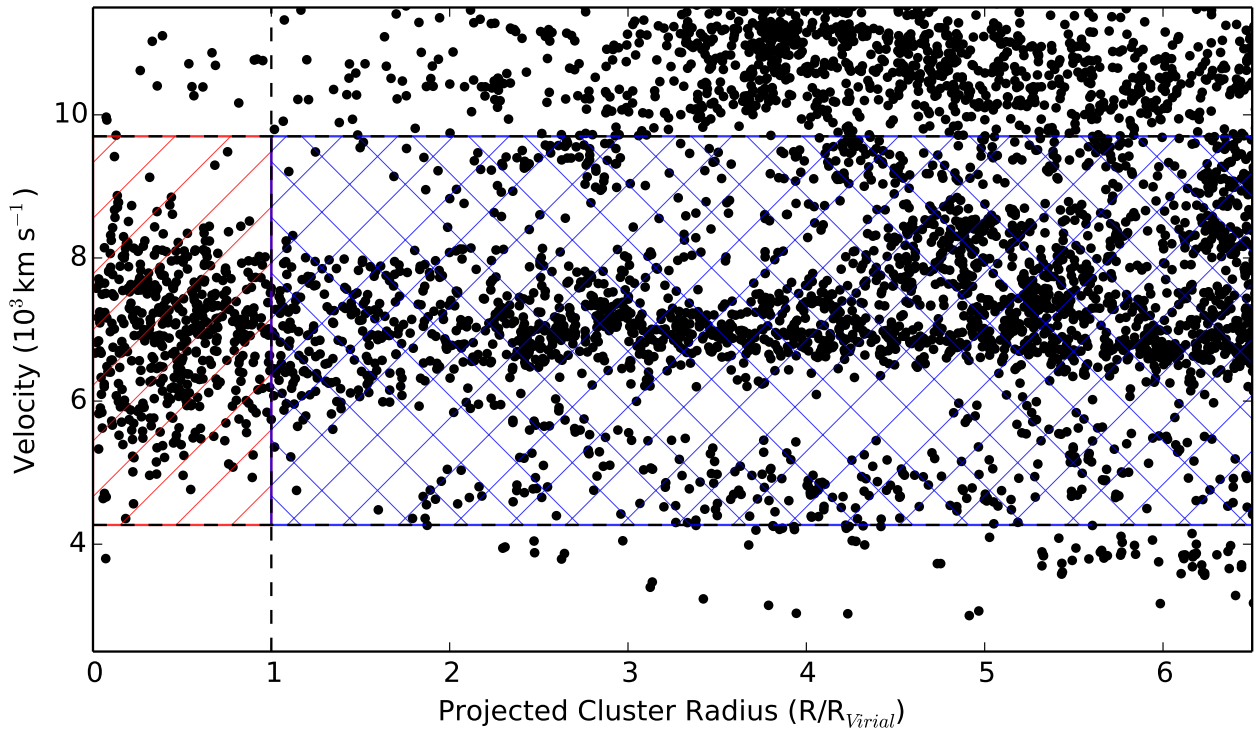}
\caption{The velocities of galaxies in the Coma cluster region plotted against distance from the Coma cluster X-ray centre (RA(J2000)=12h59m48.7s and DEC(J2000)=+27d58m50.0s). The red diagaonally-hatched box is a visual representation of the cluster selection, and the blue cross-hatched box shows the region selected for the filament sample. The filament structure is well defined out to and beyond 6$\times$ the cluster Virial radius. There are reasonably well defined voids in front of and behind the cluster.}
\label{fig:velrad}
\end{figure*}

We define a velocity selection similar to the VCC's~\citep{vcc} i.e. using the velocity dispersion ($\sigma$) of a Gaussian function fitted to galaxies within the projected Virial radius. However, this velocity dispersion is sensitive to the initial `rough' velocity selection used (see above). In order to overcome this problem we have used an iterative method to define the cluster population. We fit a Gaussian function to the velocity data, then remove galaxies that are outside of 3$\sigma$ and then re-fit the Gaussian function until the value for $\sigma$ converges. 

Figure~\ref{fig:scematic} \&~\ref{fig:velrad} indicate that this is an appropriate method for defining the cluster galaxies. The goodness of fit of a Gaussian to the histogram of velocities gives, $\chi^{2}_{dof = 23}$ = 30.1 (where $\chi^{2}_{dof = 23} =  35.2$ is equivalent to a 95\,\% confidence interval). Both in front of and behind the cluster there are natural voids that help to define the cluster sample galaxies. Table~\ref{tab:cluster_parameters} shows the limits we have used to defined the extent of the Coma cluster. 

There are 774 SDSS galaxies spectroscopically confirmed within these limits and these galaxies are henceforth referred to as members of the Coma Cluster Catalogue (CCC). The galaxies are listed in appendix A1 of the Monthly Notices of the Royal Astronomical Society on-line material.

\begin{table}
\centering
    \caption[Derived Coma cluster parameters]{Derived parameters of the Coma cluster.}
    \begin{tabular}{cc}
       \\\hline 
	Virial Radius & 3.1 Mpc \\
        Velocity Dispersion, $\sigma$           & 905  km s$^{-1}$  \\ 
        Mean Cluster velocity, $\mu$            & 6984 km s$^{-1}$ \\ 
        Minimum velocity limit, $\mu - 3\sigma$ & 4268 km s$^{-1}$ \\ 
        Maximum velocity limit, $\mu + 3\sigma$ & 9700 km s$^{-1}$ \\ \hline
    \end{tabular}
    \label{tab:cluster_parameters}
\end{table}
 
\subsection{Defining the filament}

The Coma cluster sits within a filamentary structure, which is an over density of galaxies that connects it with numerous other groups and clusters.  Defining a `filament' sample is a rather subjective process because unlike the Coma cluster it is clearly a non-Virialised structure. However, within the filament there are numerous loose-groupings and a range of galaxy densities, making it an interesting region to compare with the cluster sample. Figure~\ref{fig:velrad} clearly shows that the filament that the cluster sits within, the over density at $\sim7000$ km s$^{-1}$, extends well beyond 18 Mpc ($\sim6R_{Virial}$). These are galaxies at about the same recessional velocity as Coma, but spread in a linear structure across the sky. 
We define the filament sample as any galaxy that falls within the bounds of the NGP survey area, but outside the Viral radius of Coma. In addition each selected galaxy must satisfy the same velocity criteria i.e. $4268 < v <9700$ km s$^{-1}$ as our Coma cluster sample. This yields 951 filament galaxies, which we will refer to as the Coma Filament Catalogue (CFC). The galaxies are listed in appendix B1 of the of the Monthly Notices of the Royal Astronomical Society on-line material.

\subsection{The $Herschel$ data}

\begin{figure*}
\vspace{-5.0cm}
\centering
\includegraphics[width=\linewidth, trim =  0 0 0 0]{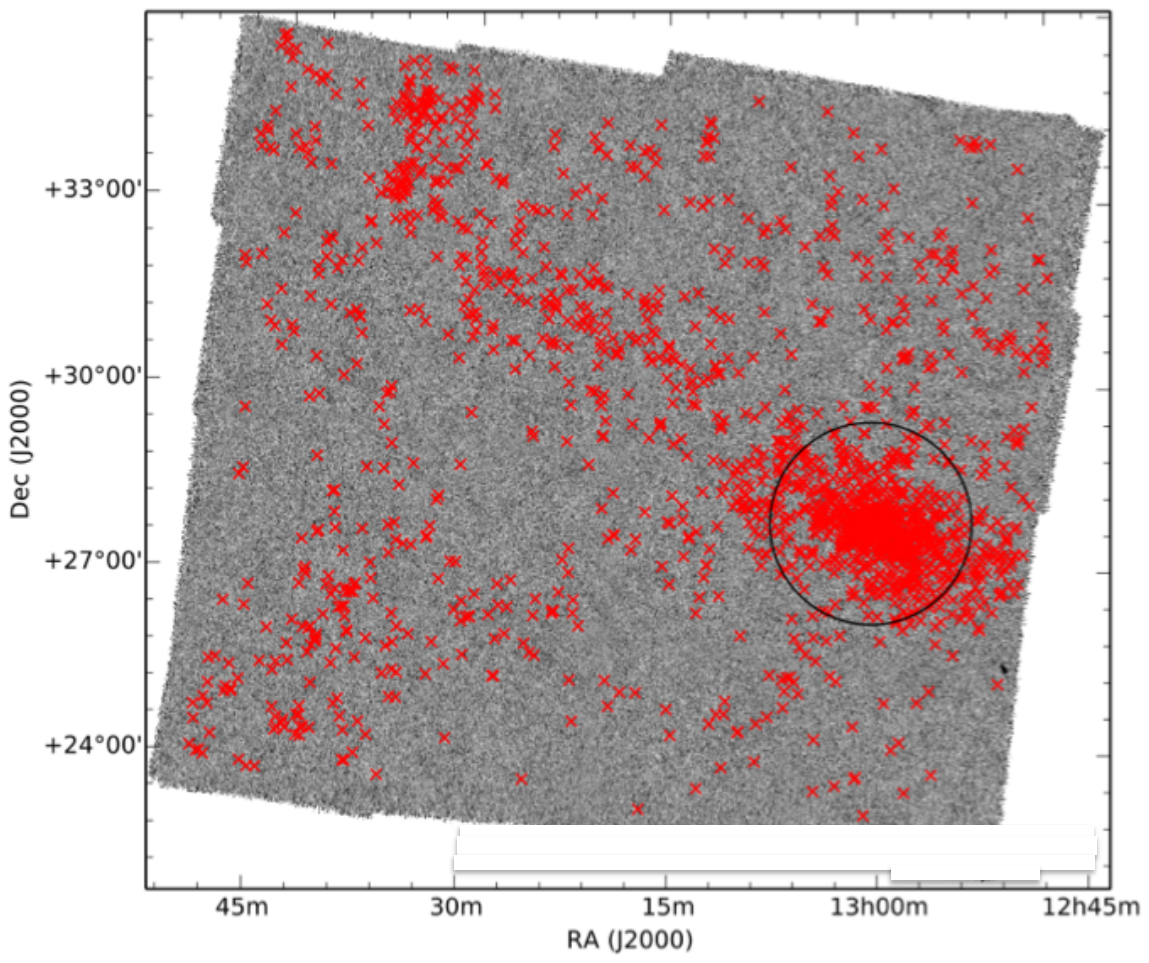}
\vspace{-6.0cm}
\caption[The NGP 250\,$\mu$m image]{The area of sky covered by the H-ATLAS NGP map. The red crosses mark the position of galaxies in both our cluster (CCC) and filament (CFC) samples. The black circle marks the approximate position of the 1.7$^{\circ}$ or 3.1 Mpc Virial radius of the cluster.}
\label{fig:ngpplot}
\end{figure*}

The NGP field is part of the $Herschel$ H-ATLAS survey \citep{eales10}. It covers approximately 15$^{\circ}$\,x\,10$^{\circ}$, centred at RA(J2000)=13h18m00.0s, Dec(J2000)=+29d00m00.0s. The survey field is shown in Figure~\ref{fig:ngpplot} along with the positions of the galaxies selected as part of the CCC and CFC samples. 

The H-ATLAS survey consists of parallel scan map data from the PACS instrument \citep{poglitsch10} at 100 and 160$\mu$m and the SPIRE instrument \citep{griffin10} at 250, 350 and 500$\mu$m. The PACS data were processed using an updated version of the pipeline presented in \cite{ibar10} that now scans the data to identify timeline jumps and glitches (see Valiante et al. in prep). Then the two scans were combined using the \textit{Scanamorphous} map maker~\citep{roussel13}. 
The SPIRE data were processed with a customised pipeline, which is very similar to the official pipeline with the exception that we used a method called BriGAdE (Smith et al., in preparation), in place of the standard temperature drift correction. BriGAdE effectively corrects all the bolometers for thermal drift without removing large extended structures like Galactic cirrus. The two scans were then combined using the mapping software in the standard $Herschel$ pipeline. 

The final NGP maps have pixel sizes of 3, 4, 6, 8 and 12 arc seconds and $1\sigma$ noise measured over the entire image of 0.8, 1.1, 0.9, 0.9 and 1.1\,mJy\,pixel$^{-1}$ (or 24.0, 43.8, 11.3, 11.6 and 12.4 mJy\,beam$^{-1}$) for 100, 160, 250, 350 and 500\,$\mu$m, respectively. Beam sizes are of order 2-3 times the pixel size so at the distance of Coma, where 10 arc seconds \,$\simeq$\,5\, kpc, we have the potential to crudely resolve some of the larger galaxies. For example, the two biggest galaxies in the cluster are NGC\,4839 and 4889 with optical diameters ($D_{25}$) of 3.6 and 3.3 arc minutes, respectively.  

\subsection{Comparsion with HeViCS and HeFoCS}
\label{sec:comp}

Below we will compare the Coma Herschel data with that previously published by us on the Virgo \citep{Davies14} and Fornax \citep{Fuller14} clusters. These data were obtained as part of the Herschel Virgo Cluster Survey (HeViCS) and the Herschel Fornax Cluster Survey (HeFoCS). The FIR maps of all three surveys were created using identical data reduction techniques. However, they differ with respect to depth and spatial coverage of the clusters.  

First, we consider the depth of the three surveys. The HeViCS maps consist of 8 scans (4 pairs of orthogonal cross-linking scans), which is two and four times as many scans as the HeFoCS and H-ATLAS data respectively. For detector random Poisson noise we would expect $\sim$\,$\sqrt{2}$ and $\sim$\,$\sqrt{4}$ increase in instrumental noise in the latter two surveys compared to HeViCS. However, ~\citet{auld12} calculated separately the instrumental and confusion noise and showed that the HeViCS SPIRE bands were effectively confusion noise limited (70\,\% of the overall noise is from confusion at 250\,$\mu$m). Consequently, when planning the HeFoCS survey only  four scans were requested as this produced confusion limited maps with half the time required for a single HeViCS tile. The H-ATLAS (NGP) survey was designed to cover as much area as possible and consequently with its two scans does not reach the confusion noise at 250$\mu$m.  

In order to measure the ratio of the global noise in the HeViCS, HeFoCS and H-ATLAS NGP maps, we measure the pixel-pixel fluctuations and apply an iterative 3$\sigma$ clip to remove bright sources. The measured `global' noise in the HeViCS, HeFoCS and NGP at 250\,$\mu$m is: 7.5, 8.9 and 11.3 mJy\,beam$^{-1}$, respectively. Yielding measured noise ratios between the HeViCS-HeFoCS and HeViCS-NGP of 1.2 and 1.5, respectively i.e. not simple Poisson noise because of the constant contribution of source confusion - thus the surveys are better suited for comparison than initial expectation.

Second, we consider the coverage of the HeViCS, HeFoCS and NGP FIR maps of their respective clusters. The Coma cluster is covered well beyond the Virial radius, however calculating the coverage for the other surveys is less straightforward. The clusters have very different physical sizes and states of relaxation - Virgo is far more `clumpy' than Fornax or Coma \citep{Davies14}. The HeViCS FIR maps consist of four tiles each of $4\,^{\circ}\,\times\,4\,^{\circ}$ running North to South \citep{davies12a}, whereas the HeFoCS is only a single tile of $4\,^{\circ}\,\times\,4\,^{\circ}$ \citep{Fuller14}. Virgo and Fornax are at about the same distance from us ($\sim$17 Mpc), but Fornax is physically much smaller than Virgo (by about a factor of four). However, the Herschel data does include about the same fraction of VCC and FCC galaxies - about $\frac{2}{3}$ of them in both cases. Of course more problematic when comparing the properties of the clusters is the more than five times greater distance to Coma than to either Virgo or Fornax - this and the relevant areas covered by the three surveys will be discussed further below.

\subsection{Flux Measurement}

\subsubsection{General approach}

\begin{figure*}
\centering
\includegraphics[trim = 35 0 45 15, clip,width=\linewidth]{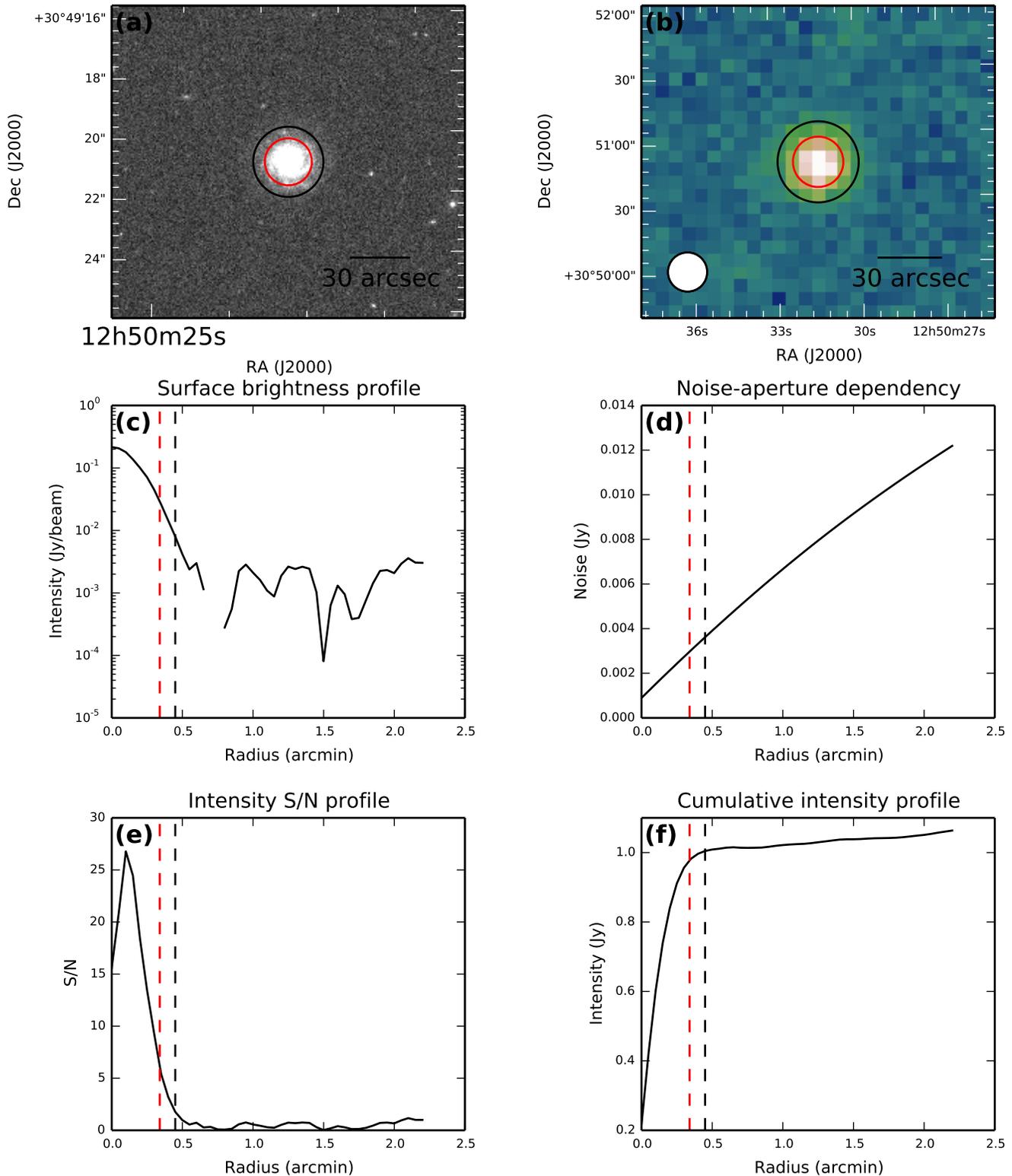}
\caption[Diagnostic output of source measurement program]{The above shows an example of the output generated for each detected CCC and CFC galaxy after we have carried out our automated photometry. Excluding the upper left hand panel, all panels refer to the $Herchel$ 250\,$\mu$m band. (a) An optical image of the galaxy, the red and black ellipses show the optical $D_{25}$ and the FIR extent of the galaxy (see text for definition). (b) The raw sub-image cutout of the FIR map. The beam size is shown in the lower left hand corner. (c) The surface brightness profile. (d) Noise for an equivalent sized circular aperture (see text for definition). (e) S/N per annuli. This shows the cut off when S/N$\le$2. (f) A cumulative intensity profile. The red and black dashed lines show the optical and FIR extent, respectively.}
\label{fig:fccoutput}
\end{figure*}

We have used a semi-automated flux measurement program, to measure the FIR flux density of each galaxy in the CCC and CFC. This program is fully described and extensively tested by~\citet{auld12}. The method is briefly described below. 

The optical parameters (position, eccentricity, optical diameter $D_{25}$ and position angle) from the optical catalogues were used to make an initial estimate of the shape and size of the FIR emission. Previous studies~\citep{cortese10, pohlen10} show that FIR emission is well traced by the optical parameters of late-type galaxies. Whereas early-type galaxies typically show more compact dust emission~\citep{smith12}. The optical parameters are only used to make an initial estimate for creating masks. The program then iterates, to create new masks and apertures (of the same size for each FIR band) that best match the diameter and ellipticity of the FIR emission. The following explanation of the process relates to the information shown in Figure~\ref{fig:fccoutput}. 

The flux measurement process starts by extracting a 200\,$\times$\,200 pixel sub-image from the raw map as shown in Figure~\ref{fig:fccoutput}b. To measure the background of the sub-image, all nearby galaxies including the galaxy being measured are initially masked at 1.5\,$\times$\,D$_{25}$. If the optical extent of the galaxy is such that this sub-image is not large enough to give an accurate background estimation, then the program will increase the size of the sub-image, up to 600$\times$600 pixels for SPIRE and 1200$\times$1200 pixels for PACS data. 

The background estimation has to deal with the near confusion limited SPIRE maps and instrumental noise in the PACS maps. The program was originally written for use with the HeViCS maps where Galactic cirrus was also a major problem. In order to remove bright background galaxies and galactic cirrus~\citet{auld12} used a 98\% flux clip and then fitted the remaining pixels with a 2D polynomial. The flux clip, removes bright background galaxies by masking out the brightest 2\% of pixels, this ensures that the 2D polynomial is only fitting the background. Unlike the Virgo fields cirrus is not obviously present in the NGP maps and so the 98\% clip is retained, but  then we just use the median pixel value of the masked sub-image as the background value.  

For each detected source we measure total flux, surface brightness, aperture noise (fully described in Section~\ref{sec:noise}) and signal to noise (S/N) around annuli of increasing radii centred on the galaxy's optical centre. The shape of the annuli is based on the galaxy's optical parameters convolved with the appropriate point spread function (PSF). As an example we plot the corresponding radial profiles in Figure~\ref{fig:fccoutput} c, e \& f, respectively. The FIR diameter\footnote{As many galaxies are not resolved $D_{FIR}$ in this case will be defined by the PSF of the $Herschel$ beam and will not be representative of the extend of dust in the galaxy.} $D_{FIR}$ is defined where the S/N profile drops below 2. This $D_{FIR}$ is used to replace the 1.5$\times D_{25}$ used to make the initial mask. The process iterates until the mask and the $D_{FIR}$ value converge. We then applied an aperture correction as given in ~\citet{ibar10} - this takes account of the encircled energy fraction within the chosen aperture size. 

For objects where the total S/N value was less than 3, the sub-image was searched for a point source. After convolving with the relevant wavelength dependent point spread function (PSF), the maximum pixel value within the FWHM of the PSF centred on the optical position was taken as the flux. The noise was calculated according to~\citet{marsden09} and~\citet{chapin11}. This involves plotting a histogram of all the pixel values in the PSF-convolved sub-image and fitting a Gaussian function to the negative tail. The FWHM of this Gaussian is then used to estimate the combined instrumental and confusion noise. This has been summed in quadrature with the calibration uncertainty (see below) to obtain a value for the total noise. If the S/N was still less than 3 we consider the object undetected and set an upper limit on the flux equal to 3 times the noise in the PSF-convolved sub-image. So, for the optically selected galaxies that do not have a FIR detection we have set an upper limit on their FIR flux density equal to the 3$\sigma$ noise from the PSF-convolved map.
This marked the end of the automatic source measurement process. The output is in the form of data files for each galaxy and the visual output shown in Figure~\ref{fig:fccoutput}.  Tables A1 and B1 in the appendix lists flux measurements (or lack of) for all galaxies in the CCC and CFC samples (see the Monthly Notices of the Royal Astronomical Society on-line material).

\subsubsection{Total error estimate}
\label{sec:noise}

The total error on each flux density measurement is estimated from the calibration uncertainty, $\sigma_{cal}$ and aperture uncertainty, $\sigma_{aper}$, summed in quadrature.

For SPIRE, $\sigma_{cal}$ is based on single scans of Neptune and on an assumed model of its emission. The final error for each band is estimated to include 4\% correlated and 1.5\% from random variation in repeated measurements. The SPIRE observer's manual\footnote{http://herschel.esac.esa.int/Docs/SPIRE/html/spire\_om.html} suggests that these should be added together, leading to a SPIRE $\sigma_{cal}$ of 5.5\%. 

For PACS, $\sigma_{cal}$ is based on multiple sources with different models of emission. The PACS observer's manual\footnote{http://herschel.esac.esa.int/Docs/PACS/html/pacs\_om.html} lists the uncorrelated uncertainties as 3\,\% \& 4\,\% for 100 and 160\,$\mu$m, respectively, and the correlated uncertainty is given for point sources as 2.2\,\%. However, the data used for calculating these uncertainties were reduced and analysed in a different way to our data. ~\citet{auld12} considered this problem and concluded that there was about a 12\,\% calibration uncertainty, which is the value we adopt here.

To calculate the aperture uncertainty ($\sigma_{aper}$) a large number of apertures of a fixed size were placed randomly on each sub-image. We measure the total flux in each aperture, then by applying an iterative 3$\sigma$ clipping procedure use $\sigma$ as the uncertainty for that aperture. Repeating this for a range of aperture sizes allows us to estimate the aperture uncertainty as a function of size. This method takes into account, both confusion noise and instrumental noise. Figure~\ref{fig:fccoutput}d shows a plot of aperture noise against radial distance.

\subsubsection{Dealing with Blending and Contamination}
$Herschel$'s comparatively large beam size can lead to unavoidable contamination by FIR background sources, which could be falsely identified with the optical source. Due to the distance of Coma many of the CCC and CFC galaxies are point rather than extended detections in the $Herschel$ data. Thus  `by-eye' inspection to reject background sources is more ambiguous than it was for the HeViCS and HeFoCS. However, the comparatively large survey area allows us to use a `Monte-Carlo' method to model the expected contamination for a given FIR-optical source separation.  
We will assume that if extended emission is detected at the location of an optical Coma galaxy it is a reliable detection. Consequently, the following discussion only applies to the point-source population, which is $\sim$60\,$\%$ of the total detections at 250\,$\mu$m.  

\begin{figure}
\centering
\includegraphics[width=\linewidth]{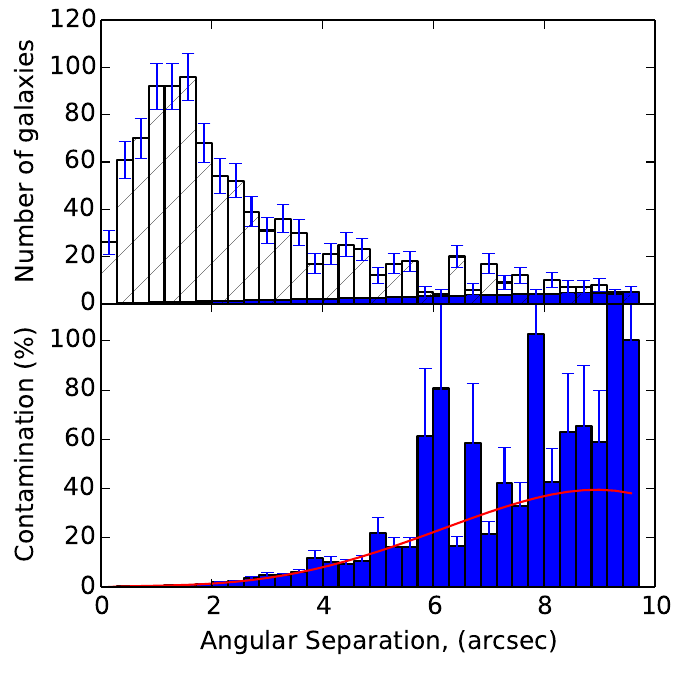}
\caption[Angular contamination simulation]{The above shows the results of our simulations and our method for estimating possible contamination by background sources at 250\,$\mu$m. The clear hatched histogram in the upper panel shows the distribution of angular separations of optically detected CCC/CFC galaxies from their nearest FIR detected neighbour. The blue histogram in the upper panel (most easily seen in the bottom right hand corner) shows the mean distribution of our simulations of a random catalogue of galaxies. The lower panel shows the contamination for each bin of angular separation. The red line is a third order polynomial fitted to the data using a $\chi^{2}$ minimisation technique.}
\label{fig:angular_contamination}
\end{figure}

As was done for HeViCS and HeFoCS we impose a strict criterion that a galaxy must be detected at 250\,$\mu$m to be included as a Herschel FIR detection - 250 $\mu$m provides the best combination of sensitivity and resolution (see below). Thus, we estimate the contamination at 250\,$\mu$m. We have done this by comparing the measured separation between an optically detected galaxy in our catalogue and a FIR detected source, with that for a random catalogue of equal size. 

We have computed the distance from the optical source to the nearest FIR neighbour using the North Galactic Point Source Catalogue (NGSPSC)\footnote{The NGPSC is a list of $Herschel$ 250$\mu$m detections in the NGP field (Valiante et al. in preparation).}. The white diagonally hatched histogram in the upper panel of Figure~\ref{fig:angular_contamination} shows the distribution of nearest neighbour separations. Clearly quite a lot of CCC/CFC optically selected galaxies have a FIR source within $\sim$2 arc sec. We have then repeat this process, using a catalogue of equal length to the CCC/CFC, but with random positions on the sky. In order to minimise the error in this latter step we repeat this random catalogue generation and cross-match many ($\sim 10^{6}$) times. The random-NGPSC separation distance distribution is shown as a blue histogram in the upper panel of Figure~\ref{fig:angular_contamination} - it is a little difficult to see because the numbers are small over these angular separations, but this does indicate that contamination is not a major problem - the association of optical and FIR sources is clearly very different if the optical galaxies are laid down at random. Combining the above two results, the lower panel of Figure~\ref{fig:angular_contamination} shows the percentage contamination in our catalogue for each bin of angular separation. We have then fitted a third order polynomial using a $\chi^{2}$ minimisation technique and calculated the angular separation where the contamination is equal to 5\,\%, we find this to be at $\sim4$ arc seconds. Given that our pixel size at 250 $\mu$m is 6 arc sec we are essentially requiring that the far infrared source position corresponds with the optical source position to plus or minus one 250 $\mu$m pixel. 

\subsection{Flux Verification}
In order to verify our automated FIR source measurement process we cross match and compare our measured fluxes with FIR data from the literature.
\begin{figure}
\centering
\includegraphics[width=\linewidth]{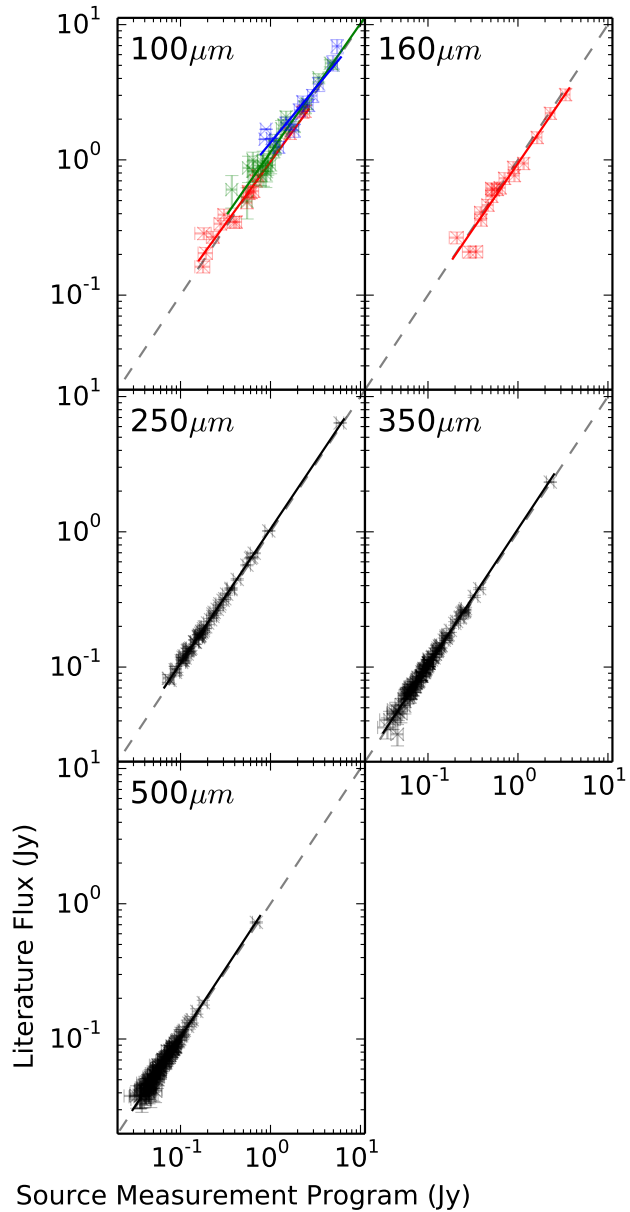}
\caption[FIR flux verfication for the NGP-CCC/CFC galaxies]{The FIR CCC/CFC fluxes plotted against other measured values. Markers are as follows; 
IRAS faint source catalogue~\citep{IRAS-FAINT}, green; 
IRAS point source catalogue~\citep{IRAS-POINT}, blue; 
deep PACS~\cite{Hickinbottom14}, red; 
and NGPSC, black. The gray diagonal dash line represents a linear exact relation i.e. $y = mx +c$, where $m = 1$ and $c = 0$). }
\label{fig:fluxv-coma}
\end{figure}
\begin{table}
\centering
  \caption[FIR flux verfication for the NGP-CCC/CFC galaxies]{The parameters of the straight line fits shown in Figure~\ref{fig:fluxv-coma}.}
  \begin{tabular}{cccc}
  \hline
  Band $\mu$m & Gradient, M & Intercept, C \\ \hline
  $100^{a}$      & 1.06\,$\pm$\,0.01    & 0.07\,$\pm$\,0.04   	\\
  $100^{b}$      & 1.09\,$\pm$\,0.08    & 0.11\,$\pm$\,0.21	\\
$100^{c}$	&0.92\,$\pm$\,0.06	& 0.03\,$\pm$\,0.02	\\
$160^{c}$	&0.95\,$\pm$\,0.08	& 0.02\,$\pm$\,0.04	\\
  $250^{d}$      & 1.07 \,$\pm$\,0.01    & -0.01\,$\pm$\,0.06  \\
  $350^{d}$      & 1.02\,$\pm$\,0.02    & -0.01\,$\pm$\,0.02   \\
  $500^{d}$      & 1.00\,$\pm$\,0.03    & 0.01\,$\pm$\,0.02    \\
\hline
\multicolumn{3}{l}{$a$ - IRAS FSC -~\cite{IRAS-FAINT}} \\
\multicolumn{3}{l}{$b$ - IRAS PSC -~\cite{IRAS-POINT}}\\
\multicolumn{3}{l}{$c$ - Deep PACS -~\cite{Hickinbottom14}} \\
\multicolumn{3}{l}{$d$ - NGPSC} \\
  \end{tabular}
  \label{tab:fluxv-coma}
\end{table}
The result is shown in Figure~\ref{fig:fluxv-coma} with gradients and intercepts tabulated in Table~\ref{tab:fluxv-coma}. We matched our catalogue with the IRAS point source catalogue~\citep{IRAS-POINT}, the IRAS faint source catalogue~\citep{IRAS-FAINT},~\citet{Hickinbottom14} a deep $Herschel$ PACS survey of the Coma cluster core, and with the NGPSC data. Table~\ref{tab:fluxv-coma} shows overall that the results are consistent with a gradient of unity and an intercept of zero. However, the faintest ($\textless 50$\,mJy) matched sources from~\citet{Hickinbottom14} appear brighter than the flux recorded at 100 and 160\,$\mu$m in our catalogue. These sources were all recorded at fluxes below the 3$\sigma$ global noise limit of the NGP PACS maps. After extensive testing by both ourselves and Hickinbottom et al. we have placed a further criterion for a source to be detected in the PACS bands i.e. we require all PACS sources to have a flux density greater than $3\times$ the global measured noise in the NGP PACS maps before we record it as a detection. Flux density or upper limit values are given for all CCC and CFC galaxies in appendices A and B of the Monthly Notices of the Royal Astronomical Society on-line material.

\section{Spectral energy distribution fitting}
\label{sec:sed-1}

As in \cite{davies12a} we have fitted a modified blackbody to every galaxy detected in all five $Herschel$ bands (198 galaxies) using the following function: 

\begin{equation}
 S_{\lambda } = \frac{\kappa_{abs}(\lambda) M_{dust} B(\lambda , T_{dust})}{D^{2}}
\end{equation}

\noindent where $S_\lambda$ is the flux density at wavelength $\lambda$, $M_{dust}$ is the dust mass, $T_{dust}$ is the dust temperature, $B(\lambda , T_{dust})$ is the Planck function, D is the distance to the cluster (105 Mpc) and $\kappa _{abs}$ is the dust absorption coefficient. The latter follows a power law modified by an emissivity ($\beta$), such that:

\begin{equation}
 \kappa_{abs}(\lambda) =  \kappa_{abs}(\lambda_{0}) \times \left ( \frac{\lambda_{0}}{\lambda}\right ) ^{\beta} 
\end{equation}

We assume that emission at these wavelengths is purely thermal and from dust at a single temperature with a fixed $\beta = 2$ emissivity \citep{davies12a}. We use $\kappa_{abs}(350 \,\mu$m) = 0.192\,m$^{2}$\,kg$^{-1}$ according to~\citet{draine03}. The above function is fitted using a $\chi^{2}$ minimisation technique. 

Although this is most likely an overly simplistic analysis, this approach has been used in previous works~\citep{davies10, davies12a, smith12, auld12, Verstappen13} and shown to fit the data very well in the FIR/sub-mm regime.~\citet{bianchi13} showed that using a single component modified blackbody returns equivalent results to more complex models such as those described in ~\citet{draine07}. From our fits we obtain dust masses and temperatures. Spectral energy distribution data and modified blackbody fits are shown in Appendix D and 250 $\mu$m images in Appendix C of the Monthly Notices of the Royal Astronomical Society on-line material.

\section{Auxiliary Data}
In this section we describe additional available data that we can make use of in our subsequent analysis.  As described above the H-ATLAS NGP region is covered by the SDSS spectroscopic and photometric surveys, which provides not only the positions, magnitudes, sizes and shapes as used above, but also star formation rates (SFR), stellar mass and metallicity for a significant fraction of our detected objects. We will also take advantage of available atomic gas data. 

\subsection{Stellar Mass, Star Formation Rate and Metallicity}
\label{sec:stellar_mass_sfr_sdss}

\citet{Brinchmann04} (see also \citet{Kauffmann03} and \citet{tremonti04} who have carried out photometry and spectral analysis respectively for the same sample) have used SDSS spectra and optical colours to calculate stellar mass\footnote{In order to make the stellar masses in Coma consistent with our previous Fornax  \citep{Fuller14} and Virgo \citep{Davies14} papers we have adjusted the SDSS stellar masses by +0.15\,dex to converted from the Kroupa IMF~\citep{Kroupa02} to a `diet' Salpeter IMF~\citep{bell03} as per the recipes in~\citet{Brinchmann04}. }, SFR and gas phase metallicities for galaxies that reside in the NGP area. Below we will use this data as part of our analysis of the galaxies in our FIR sample. \citet{Brinchmann04} designate a galaxy as either having emission lines or not. For galaxies without emission lines stellar masses and SFRs are calculated using the optical colours and the 4000\,$\AA$ break in the SDSS spectra. However, galaxies without emission lines do not have measured metallicities. In Table~\ref{tab:N_samples} we indicate the number of galaxies in the CCC/CFC with and without measured metallicities - morphologies are defined in section 4.3.

\subsection{Atomic Hydrogen}
\label{sec:atomic}

We have cross-matched the CCC/CFC with data collected by~\citet{gavazzi06b} and for galaxies in the Arecibo Legacy Fast Arecibo L-band Feed Array Survey (ALFALFA)~\citep{Haynes11}.~\citet{gavazzi06b} used the Arecibo radio telescope to observe 35 spiral galaxies in the Coma supercluster. They also added all data then available in the literature, giving 92 galaxies in total. The ALFALFA 0.4$\alpha$ data~\citep{Haynes11} adds a further 138 galaxies to this. So, in total we have atomic hydrogen data for 230 CCC/CFC galaxies. Where we have a measurement from both the literature and from ALFALFA we have taken the value from ALFALFA 0.4$\alpha$. In Table~\ref{tab:N_samples} we give the numbers of galaxies detected in HI of each morphological type (see below) in the cluster and filament.

\subsection{Morphology}
The larger distance of the Coma cluster makes morphological classification far more ambiguous, especially for the fainter members of the CCC/CFC, than that for both Virgo and Fornax galaxies. The Galaxy Zoo project~\citep{galaxyzoo} covers the majority of the SDSS DR7 galaxies that are included in the spectroscopic sample. They invite members of the general public to decide if a galaxy is either an elliptical or spiral. Based on these votes each galaxy can then be assigned a probability of being elliptical $p(E)$ or spiral $p(S)$. This allows us to define 3 morphological categories; early, $p(E) \textgreater 0.8$; late, $p(S) \textgreater 0.8$ and uncertain where $p(E) \textless 0.8$ and $p(S) \textless 0.8$. We will discuss the significance of these morphological categories below. 

The \emph{GOLDMINE} database \citep{gavazzi2003} provides multi-wavelength data from a number of sources, with varying levels of completeness. As part of the \emph{GOLDMINE} database~\citet{Gavazzi96} catalogued galaxies in the Coma region. They are complete for galaxies in the B band brighter than $m_{B} = 15.5$, covering a large - although not total - fraction of the NGP survey area. They visually classify galaxy morphologies into the more familiar Hubble types (E through to Sd). However the brighter magnitude limit and reduced area coverage means that only 256 (15\,\%) galaxies are in both \emph{GOLDMINE} and our CCC/CFC catalogues. 

\begin{figure}
\centering
\includegraphics[width=\linewidth]{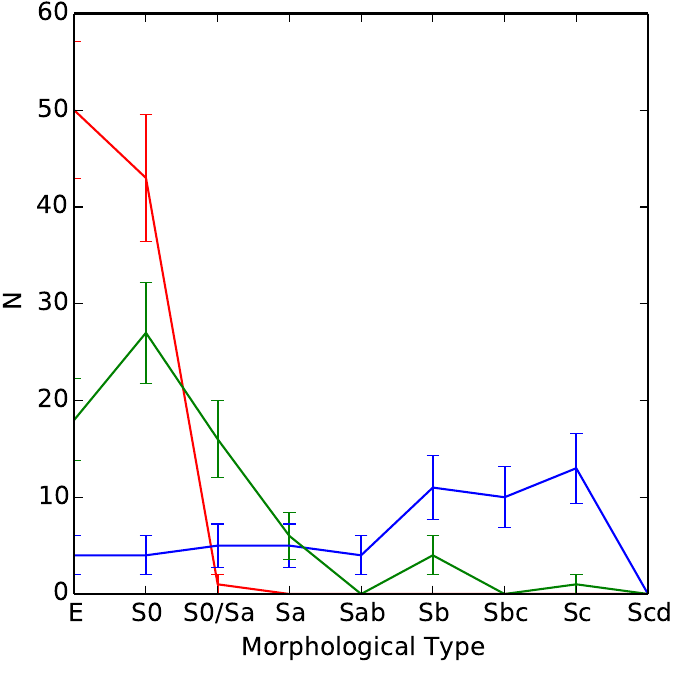}
\caption[Galaxy Zoo morphology compared to hubble type]{Histograms of morphology assignments from the \emph{GOLDMINE} database (x-axis) in each of our morphological groups from the Galaxy Zoo: early-type (red), uncertain (green) and late-type (blue).The error bars are simply root N.}
\label{fig:zoogoldmine}
\end{figure}

\begin{figure*}
\centering
\includegraphics[width=4.3cm,height=4.3cm]{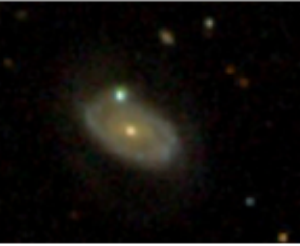} 
\includegraphics[width=4.3cm,height=4.3cm]{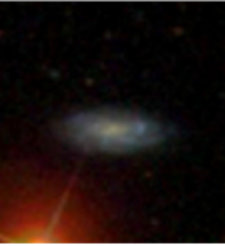} 
\includegraphics[width=4.3cm,height=4.3cm]{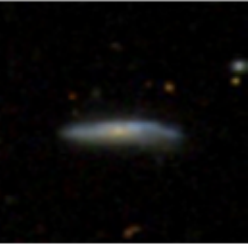} 
\includegraphics[width=4.3cm,height=4.3cm]{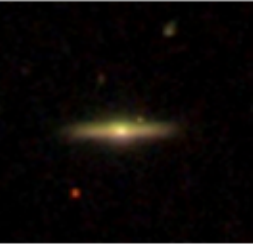} \\
\includegraphics[width=4.3cm,height=4.3cm]{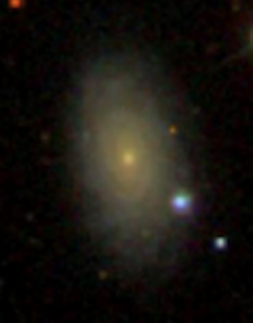} 
\includegraphics[width=4.3cm,height=4.3cm]{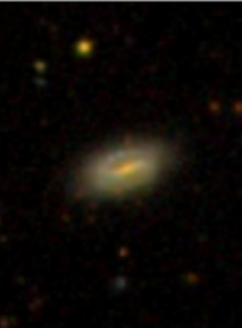} 
\includegraphics[width=4.3cm,height=4.3cm]{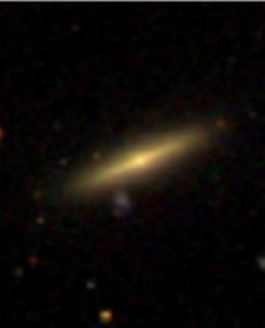} 
\includegraphics[width=4.3cm,height=4.3cm]{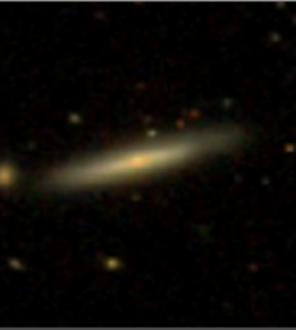}
\caption[Goldmine miss claffications]{SDSS images of eight galaxies that are classfied as early in the \emph{GOLDMINE} database, however, Galaxy Zoo gives them a greater than 80\,\% likelyhood of being a late-type. The upper row are classified as E, and the lower are classified as S0. The galaxies are sorted left-to-right with galaxies on the right being closer to edge-on.}
\label{fig:zoo_VS_goldmine}
\end{figure*}

\begin{figure*}
\vspace{-6.5cm}
\centering
\includegraphics[width=\linewidth]{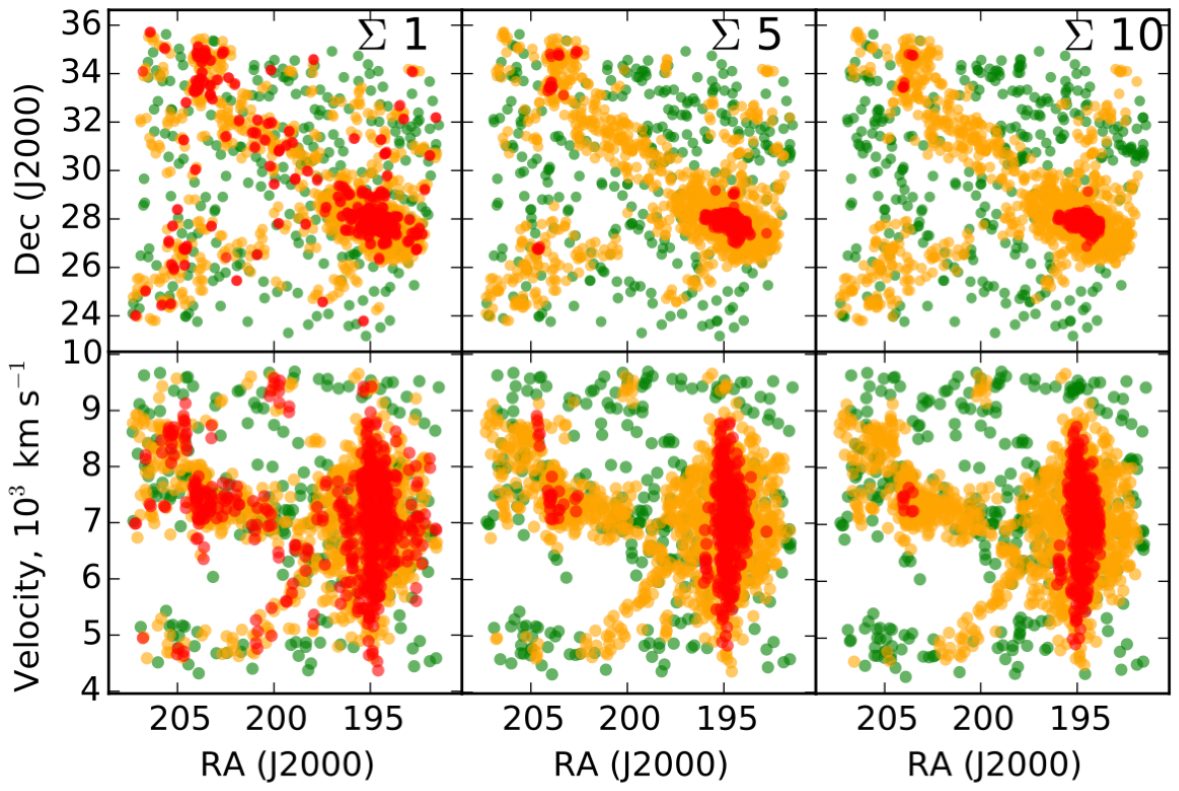}
\vspace{-7.5cm}
\caption[Local enviroment statistics of the CCC/CFC]{The above illustrates the changing values of $\Sigma_{N}$ for the $1^{st}$, $5^{th}$ and $10^{th}$ nearest neighbours in our sample (left to right). The upper three plots are spatial plots in RA and Dec, the lower three are for velocity and RA. Each point has been colour coded with green ($\Sigma_{N}<1$ Mpc$^{-2}$), amber ($1 <\Sigma_{N}<20 $ Mpc$^{-2}$), and red ($20<\Sigma_{N}$ Mpc$^{-2}$) to indicate low, medium, and higher density regions. }
\label{fig:sigma_Def}
\end{figure*}

\begin{table*}
\caption[Number of galaxies in each sample]{The tables below shows the availability of metalicity and HI data for the CCC/CFC galaxies split by morphological type (see text for details). The numbers in brackets refer to the percentage of that galaxy for a given morphological type.}
\centering
\begin{tabular}{ccccc}
\hline
Sample & N$_{late}$ & N$_{uncert.}$ & N$_{early}$ & N$_{total}$ \\ 
\hline
CCC/CFC & 474  & 963  & 288  & 1725  \\ 
CCC/CFC (+Metallicity) & 285 (60\,\%) & 295 (30\,\%) & 5 (1\,\%) & 585 (33\,\%) \\
CCC/CFC (+HI) & 119 (25\,\%) & 81 (8\,\%) & 8 (2\,\%) & 208 (12\,\%) \\
\hline
\end{tabular}
\centering
\begin{tabular}{ccccccc}
\\\\
\hline
Sample & \multicolumn{3}{c}{Cluster} & \multicolumn{3}{c}{Filament}\\
~ & N$_{late}$ & N$_{uncert.}$ & N$_{early}$ & N$_{late}$ & N$_{uncert.}$ & N$_{early}$ \\ 
\hline
CCC/CFC & 83& 504&187&391& 459&101\\ 
CCC/CFC (+Metallicity) & 25 (30\,\%)& 53 (11\,\%)&1 ($<$1\,\%)&260 (67\,\%)& 242 (53\,\%)&4 (4\,\%)\\
CCC/CFC (+HI) & 24 (29\,\%)& 21 (4\,\%)&7 (4\,\%)&95 (24\,\%)& 60 (13\,\%)&1 (1\,\%)\\
\hline
\end{tabular}
\label{tab:N_samples}
\end{table*}

The Galaxy Zoo catalogue has been shown to be consistent with classifications of the same galaxies by professional astronomers~\citep{galaxyzoo}. However,~\citet{galaxyzoo} also show that fainter galaxies are harder to classify and more-likely to be classified as an early-type or uncertain-type. In order to understand our three morphological groups, based upon the above selection, we compare our classification with that of \emph{GOLDMINE} - Fig.~\ref{fig:zoogoldmine}. 
  
Fig.~\ref{fig:zoogoldmine} clearly shows that at the distance of Coma, Galaxy Zoo is a good predictor of morphology when compared to \emph{GOLDMINE}. Early-types with a 0.8 likely-hood selection are mainly composed of E and S0's ($\sim$ 98\,\%), only one early-type is in the S0/Sa bin. Late-type galaxies are mainly composed of Sa to Sc's types. However, 8 galaxies are classified late-type by Galaxy Zoo and early-type by \emph{GOLDMINE}. In order to understand this we have visually inspected these galaxies (see Figure~\ref{fig:zoo_VS_goldmine}). Four of these galaxies (CCC\,232, CCC\,513, CFC\,57, and CFC\,344) are clearly edge-on galaxies, and as such morphological classification is always ambiguous. The remaining four galaxies are clearly late-types and so appear to be miss-classified in \emph{GOLDMINE}. 

Figure~\ref{fig:zoogoldmine} also helps us understand the morphological make up of the uncertain-type. The uncertain-type covers a range of morphologies from E to Sc however, it is mostly made up of S0 and S0/Sa galaxies. 

\subsection{Local density}
\label{sec:sigma_def}

In order to examine the affect of local structure or environment rather than large scale structure such as the cluster and filament, we simply use the N$^{th}$ nearest neighbour statistic ($\Sigma_{N}$) as defined by the equation
\begin{equation}
\Sigma_{N} = \frac{N}{\pi D_{N}^{2} }  
\end{equation}
to measure the local density of optically selected CCC/CFC galaxies. $\Sigma_{N}$ is the surface density out to the N$^{th}$ nearest neighbour that lies at a distance $D_{N}$. As we assume that all our galaxies are at $\sim$100 Mpc we effectively `collapse' our catalogues of cluster and filament galaxies into a 2D plane and measure spatial distances on the sky.

We have calculated $\Sigma_{N}$ for the $1^{st}$, $5^{th}$ and $10^{th}$ nearest neighbours for each galaxy. In order to avoid edge effects for this analysis, we have drawn our sample from a greater area than the NGP $Herschel$ field, but with our original velocity selection for the cluster and filament. 

Figure~\ref{fig:sigma_Def} illustrates the merit of each $\Sigma_{N}$ statistic and what scale of environment it traces. $\Sigma_{10}$ effectively smooths over the largest spatial scale and is sensitive to the larger scale structures. Conversely, $\Sigma_{1}$ and $\Sigma_{5}$ are sensitive to a galaxy's more immediate environment. As expected all $\Sigma_{N}$ statistics are highest in the cluster core. However, Figure~\ref{fig:sigma_Def} shows that within the filament there are a number of higher density regions that can probably be associated with small groups of galaxies. 

\section{Far-infrared results}
\subsection{Detection rates}
\label{sec:det_rates_coma}
\begin{figure}
\centering
\includegraphics[width=\linewidth]{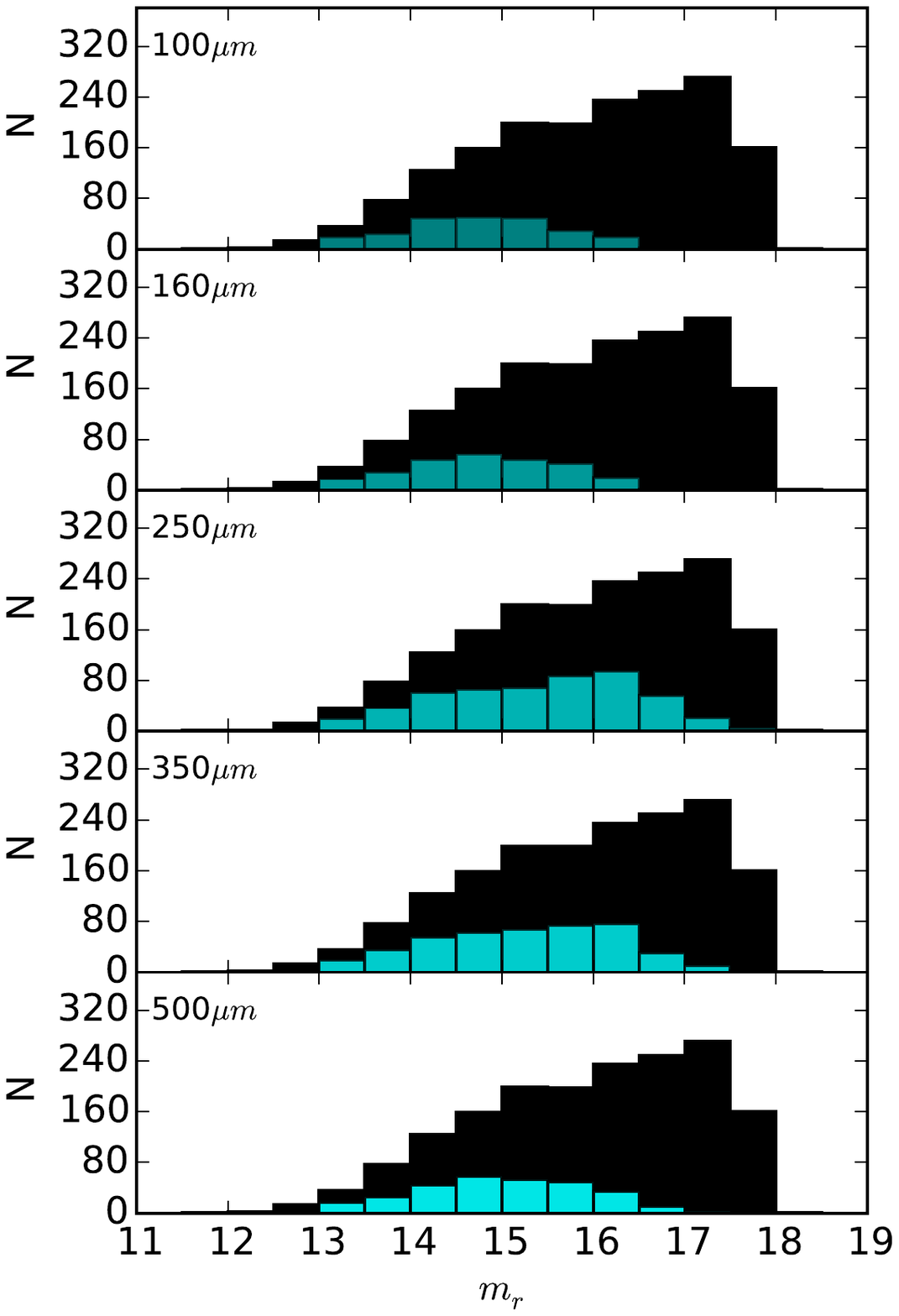}
\vspace{-1.0cm}
\caption[A histogram of optical magnitude mbt of the CCC galaxies]{A histogram of $r$ band optical magnitude ($m_{r}$) for the CCC/CFC galaxies - the black and cyan bars are the total (optical) and FIR (250 $\mu$m) detected galaxies respectively.}
\label{fig:dopt}
\end{figure}

Our Coma and filament far-infrared data is obtained from the positions of optically selected galaxies in the CCC/CFC. This is not ideal as we would prefer for our analysis to select galaxies via their far-infrared flux density. The problem of course is identifying which far-infrared sources belong to the cluster and which are in the background. Figure~\ref{fig:dopt} shows the distribution of optical magnitudes in the SDSS r band for all CCC/CFC (black) galaxies compared to just those detected at 250 $\mu$m in the Herschel data (cyan). We do not detect any galaxies in the far-infrared fainter than r = 17.5, which is below the limit of our optical selection. So, from this our expectation is that a deeper optical catalogue would not significantly increase the numbers of FIR detections in our current data. However, there is still of course the possibility of sources with unusual ratios of optical to far-infrared emission that are detected in the cluster/filament by Herschel, but are too faint to appear in the optical catalogue. 

\begin{figure}
\centering
\includegraphics[trim = 0 0 0 0,width=85mm]{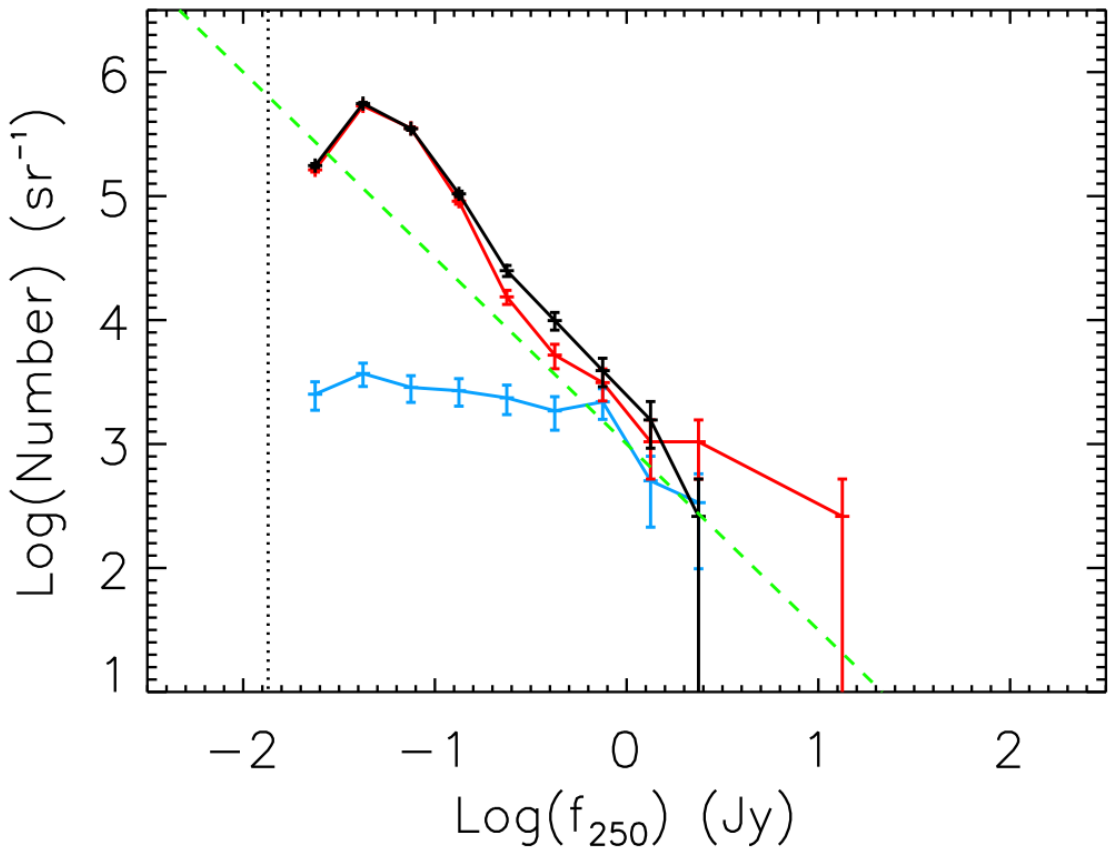}
\vspace{-3.5cm}
\caption[Histograms of FIR sources in background and forground in all 3 clusters]{The 250\,$\mu$m flux density of galaxies against number per steradian. The black line is for an eastern quadrant of the NGP field, where contamination by foreground galaxies is at its lowest (Fig~\ref{fig:ngpplot}), while the red line is for the NGP within the Virial radius of the cluster.  The blue line is for CCC galaxies detected at 250 $\mu$m. The green dashed line has a slope of 1.5 and indicates the expected counts for a non-evolving Euclidean universe. The black dotted line indicates the minimum flux density detectable in the NGP field.}
\label{fig:numbercounts}
\end{figure}

This is an issue we have investigated before for the Virgo \citep{davies12a} and Fornax clusters  \citep{Fuller14}. Our conclusion was that there is no evidence for a significant population of far-infrared sources missed by our optical selection. We have carried out a similar analysis here and come to the same conclusion. Our method is fully described in \cite{davies12a}, but briefly we compare the 250 $\mu$m number counts in an area away from the cluster with those over the cluster region, in this case within the Virial radius (see Fig~\ref{fig:numbercounts}). Although we are dealing with large numbers in the background compared to those in the cluster we find no evidence for an excess population in the cluster field. The black line in Fig~\ref{fig:numbercounts} are the counts for an eastern quadrant of the NGP field where contamination by foreground (filament) galaxies is at its lowest (Fig~\ref{fig:ngpplot}) - these are the 'background' counts. The red line shows the counts from within the cluster Virial radius, these counts only exceed the 'background' at bright flux densities due to the bright cluster sources. The counts of objects detected using our optical selection are shown in blue. There is no evidence, within the errors, for a large population of faint cluster FIR sources that are missed by having an optical rather than a FIR selection criteria i.e. the red line only exceeds the black line at bright flux densities.

With the above caveat on optical selection Table~\ref{tab:detection_rates} shows how many galaxies were detected above a 3$\sigma$ noise level in each band  in the $Herschel$ data. The SPIRE 250\,$\mu$m band has the highest detection rate, something that has previously been noted by~\citet{auld12} and~\citet{Fuller14} in Virgo and Fornax, respectively. This is due to a combination of instrument sensitivity and the typical shape of a galaxy's SED.  Consequently, we will use the 250\,$\mu$m band observations extensively in the below analysis and discussion. At 250\mic we detect 99 of 774 (13\,\%) of galaxies in the cluster and 422 of 951 (44\,\%) of galaxies in the filament - a significant difference and of course primarily due to the very different morphological mix (Table~\ref{tab:N_samples}).  

\begin{figure*}
\centering
\includegraphics[width=\linewidth]{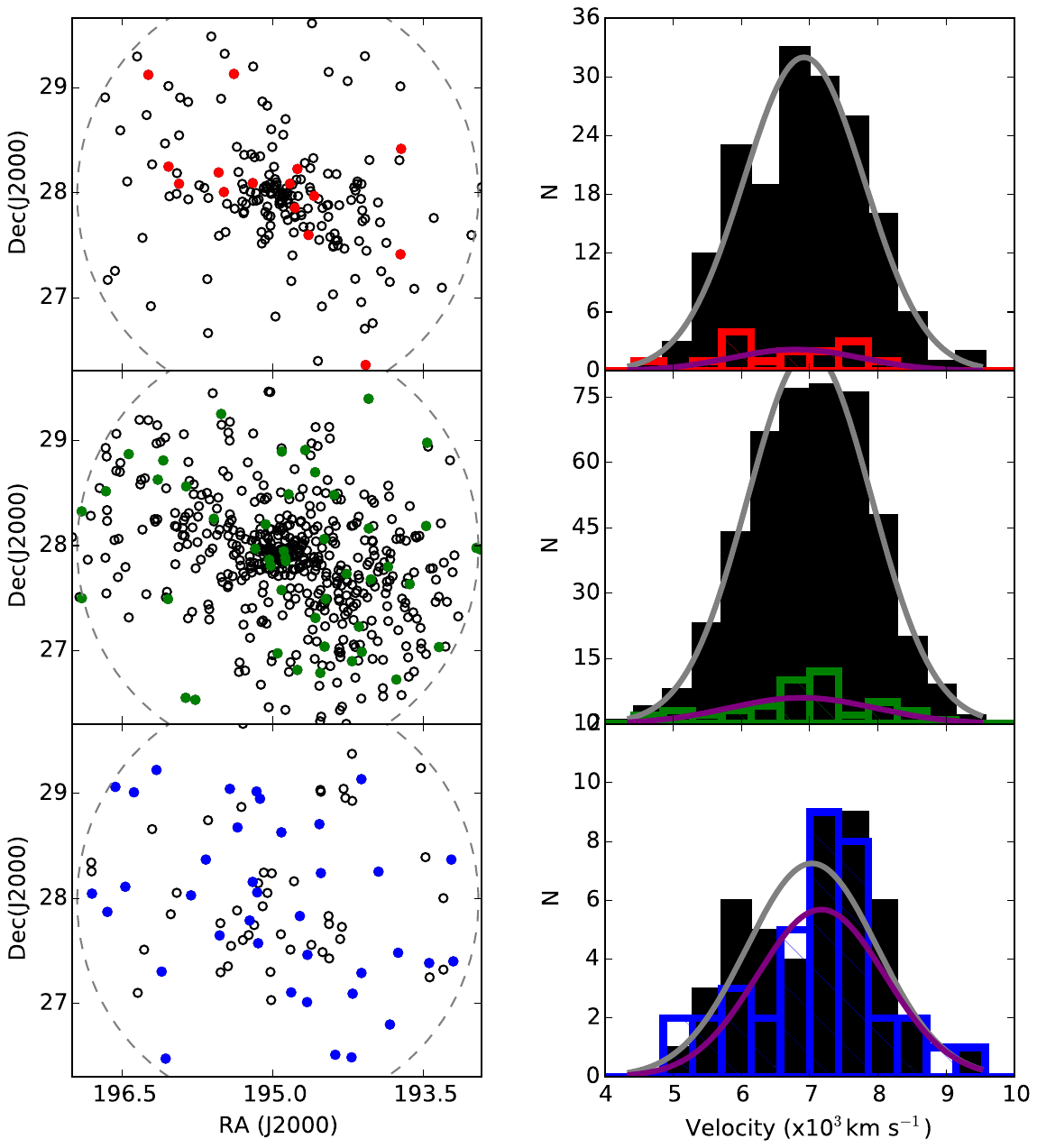}
\vspace{-3.5cm}
\caption[Location and stellar mass of detected galaxies in The CCC]{The locations and velocities of galaxies within the Virial radius of the cluster. Coloured and black markers and histograms designate far-infrared detected and undetected galaxies, respectively. The three morphological types are shown from top to bottom in red, green and blue representing early, uncertain and late-type galaxies. The left-hand panels show their location in RA and Dec. The right-hand panels show histograms of radial velocity - the purple and grey lines are Gaussian fits to the histograms of detected and undetected galaxies, respectively.}
\label{fig:cluster}
\end{figure*}

In order to better understand the properties of the galaxies we detect we can estimate the limiting dust mass required for a detection at 250\,$\mu$m. The lowest detected 250\,$\mu$m flux in our FIR catalogue is $\sim$\,15\,mJy. Assuming a dust temperature of 20\,K  (see below) the corresponding limiting dust mass is $\log_{10}(M_{dust}/$M$_{\odot}) = 6.1$. If we assume that a typical late-type galaxy has a dust-to-stellar mass ratio of $\log_{10}(M_{dust}/M_{stars}) = -3$~\citep{cortese12} then we should detect all galaxies with stellar masses of $\log_{10}(M_{stars}/$M$_{\odot}) > 9.1$. According to \citet{Brinchmann04} the stellar mass limit for a
galaxy with the mass-to-light ratio appropriate for a 13.6 Gyr old stellar population with an apparent magnitude of $r=17.8$ (the magnitude limit) and at the distance of Coma is also $\log_{10}(M_{stars}/$M$_{\odot}) \approx 9.0$.

Table~\ref{tab:sum_det_rates} shows the detection rate for each morphological type (as defined above using Galaxy Zoo) within both the cluster and filament. As expected late-type galaxies have the highest detection rate in both cluster and filament (of order \,50\,\%). Conversely early-type galaxies have the lowest detection rate (of order \,10\,\%). Both early and late-type detection rates are statically identical between the two samples, whereas, uncertain-type galaxies have far lower detection rates inside the cluster environment (9\,\% and 35\,\% for the cluster and filament, respectively). Far fewer morphologically uncertain (S0/Sa?) galaxies are detected in the far-infrared in the cluster compared to the field, suggesting that their star formation has been switched off in the cluster, but that their morphology has been unaffected.

\begin{table*}
  \caption[FIR detection rates of the CCC]{The detection rates for each $Herschel$ band in the cluster and the filament. In total there are 744 and 951 optically detected galaxies in the cluster and filament, respectively.}
\begin{tabular}{ccccc}
\hline
Band   & \multicolumn{2}{c}{Cluster} & \multicolumn{2}{c}{Filament} \\
(\mic)&Number of detections &Detection rate&Number of detections&Detection rate\\
~ &(N)&(\%)&(N)&(\%)\\
\hline
100 & 60 & 8 & 187 & 20 \\
160 & 55 & 7 & 210 & 22 \\
250 & 99 & 13 & 422 & 44 \\
350 & 80 & 10 & 355 & 37 \\
500 & 54 & 7 & 247 & 26 \\
\hline
\end{tabular}
  \label{tab:detection_rates}
\end{table*}

\begin{table*}
\caption[FIR detection rates: cluster vs filament]{Detection rates for each morphological type in the cluster and filament.}
\begin{tabular}{ccccccc}
\hline
Morphological & Cluster  & Filament \\
Type & $N \pm \sqrt{N} $ &  $N \pm \sqrt{N} $ \\
\hline
late&37 $\pm$ 6 &239 $\pm$ 16 \\
uncertain&47 $\pm$ 7 &164 $\pm$ 13  \\
early&15 $\pm$ 4 &19 $\pm$ 4  \\
\hline
\end{tabular}
\label{tab:sum_det_rates}
\end{table*}

For each morphological type we have compared the positions and velocities of the far-infrared detected and undetected galaxies (Figures~\ref{fig:cluster}). The left panels black and coloured markers indicate the spatial position of each galaxy. The dash black circle in the left-hand panel marks the Virial radius of the cluster. The right-hand panels show histograms of radial velocity. For each histogram of velocity we have fitted a Gaussian using $\chi^{2}$ minimisation, where errors are $\sqrt{N}$ galaxies per bin, the purple and grey histograms represent fits to the detected and undetected galaxies respectively. We have listed the average projected cluster radius for detected and undetected galaxies in the cluster as well as the derived velocity dispersions in Table~\ref{tab:loc_smass_fits}. 

\begin{figure*}
\centering
\includegraphics[width=\linewidth]{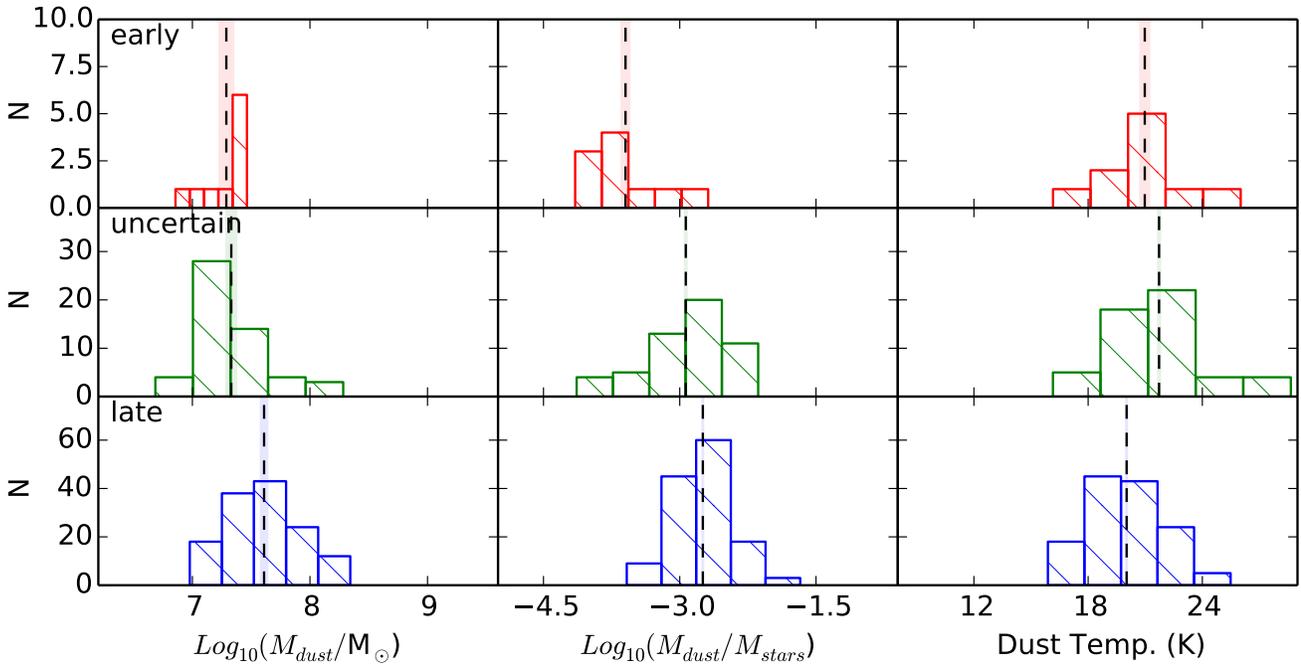}
\caption[Histograms of FIR properties]{Histograms of dust mass (left), dust-to-stellar mass (centre) and dust temperature (right) for 198 galaxies detected in all 5 $Herschel$ bands and fitted with a modified black body. Early, uncertain and late-type galaxies are plotted in red, green and blue, respectively. The vertical dashed line indicates the mean of each sample.}
\label{fig:sedresults}
\end{figure*}

\begin{table*}
\caption[Location and stellar mass of detected galaxies in The CCC]{The table below contains information relating to Figure~\ref{fig:cluster} for the cluster sample. The radius refers to the average projected cluster centric radius for either far-infrared detected or undetected galaxies of a given morphological type. The velocity dispersions have been obtained by fitting a Gaussian function to the velocity histograms.}
\begin{tabular}{ccccccc}
\hline
 Morphological Type & \multicolumn{2}{c}{Radius, (R/R$_{virial}$)} & Difference & \multicolumn{2}{c}{Velocity Dispersion, (\kms)} & Difference \\
 & Detected & Undetected & ~ & Detected & Undetected & ~ \\\hline
late & 0.48$\pm$0.04 & 0.59$\pm$0.04 & 1.9\,$\sigma$ &935$\pm$154 (0.5) & 956$\pm$117 (0.56) &0.1\,$\sigma$ \\ 
uncertain & 0.42$\pm$0.01 & 0.57$\pm$0.04& 3.6\,$\sigma$ & 1070$\pm$206 (0.9) & 906$\pm$20 (0.37) & 0.8\,$\sigma$ \\
early & 0.43$\pm$0.07 & 0.36$\pm$0.02 & 1.0\,$\sigma$& 951$\pm$206 (0.29) & 884$\pm$46 (0.67) & 0.3\,$\sigma$\\\hline
\end{tabular}
\label{tab:loc_smass_fits}
\end{table*}

\begin{table*}
\centering
\caption[A statistical comparison of cluster and filament galaxies' FIR properties]{A statistical comparison of the cluster (Sample 1) and filament (Sample 2) plus a comparison with normal and HI deficient galaxies. $\mu$ is the mean value and $\sigma$ the associated error. The K-S test compares the distribution of values in each sample. The D-value is the absolute maximum distance between the cumulative distributions of the two samples and the null hypothesis (that the two samples were drawn from the same distribution) is rejected if the $p_{value}$ is less than your acceptable significance level. }
\begin{tabular}{cccccc}
\hline
\\
Sample 1 & Sample 2 & $\mu_{1}$($\sigma_{1}$) & $\mu_{2}$($\sigma_{2}$) & \multicolumn{2}{c}{K-S test} \\
~&~&~&~&D-value& $p_{value}$
\\ \hline
\multicolumn{6}{l}{\textbf{Dust Mass (log($M_{Dust}$/$M_{\odot}$))}} \\
Cluster Early&Filament Early&7.25(0.09)&7.35(0.05)&0.333&0.89\\
Cluster Uncertain&Filament Uncertain&7.33(0.07)&7.33(0.05)&0.14&0.974\\
Cluster Late&Filament Late&7.66(0.06)&7.6(0.03)&0.12&0.914\\
HI-Normal&HI-Deficent&7.64(0.06)&7.72(0.05)&0.317&0.5\\
\\
\multicolumn{6}{l}{\textbf{ Stellar Mass / Dust Mass (log($M_{dust}/M_{stellar}$)})} \\
Cluster Early&Filament Early&-3.74(0.14)&-3.39(0.23)&0.583&0.254\\
Cluster Uncertain&Filament Uncertain&-3.05(0.14)&-2.89(0.07)&0.168&0.892\\
Cluster Late&Filament Late&-2.78(0.06)&-2.74(0.03)&0.144&0.762\\
HI-Normal&HI-Deficent&-2.98(0.11)&-2.65(0.04)&0.544&0.036\\
\\
\multicolumn{6}{l}{\textbf{Dust Temp. (K)}} \\
Cluster Early&Filament Early&22.12(0.77)&19.25(0.9)&0.833&0.03\\
Cluster Uncertain&Filament Uncertain&21.51(0.49)&21.81(0.43)&0.179&0.845\\
Cluster Late&Filament Late&20.66(0.44)&19.88(0.18)&0.229&0.205\\
HI-Normal&HI-Deficent&20.46(0.99)&20.37(0.32)&0.213&0.918\\
\hline
\end{tabular}
\label{tab:stats-coma}
\end{table*}

All three morphological types have velocity dispersions that are statistically identical for FIR detected and undetected galaxies. All the velocity dispersion are also statically identical to the overall velocity dispersion ($\sim$\,900\,\kms) of the cluster. If any of the morphological populations had a velocity dispersion that was greater than that of the cluster as a whole it would imply that it was less relaxed and thus probably far later in joining the cluster. The fact that all our morphological types show no evidence of a different velocity dispersion implies that they have been in the cluster longer than a crossing time of a few Gyr. This is longer than the gas stripping timescale even for the most massive galaxies \citep{boselli06}. 

With regard to position in the cluster, as expected late-type galaxies have the largest average projected cluster radius (0.55\,$R_{virial}$). Conversly early-type galaxies have the smallest average projected cluster radius (0.37\,$R_{virial}$). Both late and early-type galaxies show no difference in mean radius between FIR detected and undetected galaxies ($\textless 2 \sigma$). Uncertain-type galaxies are the only sample that shows any significant difference in spatial position when selected by their far-infrared properties (0.4 and 0.6$R_{virial}$ for detected and undetected galaxies respectively). This is somewhat counter intuitive, with the detected galaxies having an average spatial position closer to the cluster centre than those not detected.

\subsection{Analysis of SED fits, dust masses \& temperatures}
\label{sec:sed-fits}

There are 521 CCC/CFC galaxies detected in the far-infrared at 250 $\mu$m, but not all of these are detected in all five $Herschel$ bands.  For the 198 galaxies detected in all five bands we have fitted their SEDs with single temperature modified blackbody curves using $\beta=2$. The average $\chi^{2}$ goodness of fit value for the entire sample is $< \chi^{2}_{dof = 3} > = 2.7$ (where $\chi^{2}_{dof = 3} < 7.8$, corresponds with a confidence level of 95\,$\%$) indicating that for the majority of the sample the SEDs are well fitted. 7 galaxies have $\chi^{2}_{dof = 3} > 7.8$ indicating that a single temperature component may not be sufficient to model their FIR emission.  In appendix C (see of the Monthly Notices of the Royal Astronomical Society on-line material) we show the SEDs and modified black body fits for each galaxy. Histograms of dust mass, dust-to-stellar mass and dust temperatures produced from the SED fitting process are shown in Figure~\ref{fig:sedresults}. 

Figure~\ref{fig:sedresults} shows that detected late-type galaxies have dust masses ranging from $\log_{10}(M_{dust}$/M$_{\odot}) = 7.0$ to $8.3$ and temperatures of 16.0 to 25.5\,K, with mean values of $\log_{10}(M_{dust}$/M$_{\odot}) = 7.63 \pm 0.02$ and 20.0\,$\pm$\,0.1\,K. Uncertain-type galaxies have dust masses ranging from $\log_{10}(M_{dust}$/M$_{\odot}) = 6.7$ to $8.3$ and temperatures of 16.2 to 28.7\,K, with mean values of $\log_{10}(M_{dust}$/M$_{\odot}) = 7.33 \pm 0.04$ and 21.7\,$\pm$\,0.5\,K. Early-types have a narrower range of dust masses of $\log_{10}(M_{dust}$/M$_{\odot}) = 6.9$ to $7.5$ and temperatures of 16.2 to 26.0\,K, with mean values of $\log_{10}(M_{dust}$/M$_{\odot}) =7.3 \pm 0.04$ and 21.0\,$\pm$\,0.2\,K. This observed warmer dust temperature for early type galaxies has previously been observed in both the Virgo  \citep{Davies14} and Fornax \citep{Fuller14} clusters and may be due to  additional heating by hot x-ray gas.

CCC/CFC galaxies have mean dust-to-stellar mass ratios of $\log_{10}(M_{dust}/M_{stars})$ = -3.6\,$\pm$\,0.04, -2.93\,$\pm$\,0.01 and -2.75\,$\pm$\,0.01 for early, uncertain and late-types, respectively. As expected from our previous results for Virgo and Fornax~\citep{auld12,Fuller14}, late-types have richer and cooler dust reservoirs, and early-types have a relatively depleted and warmer ISM. 

In Table~\ref{tab:stats-coma} we use the Kolmogorov-Smirnov two sample test (K-S) to make a more quantitative comparison between the cluster and filament's early, uncertain and late-type galaxy populations with respect to dust mass, dust-to-stars ratio, and dust temperature. The K-S test shows that for a given morphological type the FIR properties of galaxies in the cluster and filament are statistically very similar with the one obvious exception of early type galaxy dust temperatures, early-types have hotter dust temperatures in the cluster environment. Just why this occurs is not clear, though it is possible it has something to do with the cluster x-ray gas. The above results suggest however, that on the whole the different environments have had very little effect on the dust properties of their constituent galaxies. 

We have also used (Table~\ref{tab:stats-coma}) \hi-defficiency values calculated for 70 galaxies from~\citet{gavazzi06b} to compare galaxies that are \hi-normal (\hi-def. $\ge$ 0.5) with \hi-deficient (\hi-def. $<$ 0.5) - these galaxies are exclusively late-type. A K-S test shows that \hi-normal and \hi-deficient galaxies are statically identical with respect to dust mass and temperature. However, with respect to dust-to-stellar mass ratio \hi-normal and \hi-deficient galaxies are unlikely to be drawn from the same distribution. Galaxies that are depleted in HI are also depleted in dust \citep{cortese12}. ~\citet{gavazzi06b} has previously showed that \hi-deficiency decreases with cluster centric distance with the most \hi-deficient galaxies at the cluster centre. The implication is that \hi-deficiency traces those galaxies most strongly affected by the ram pressure stripping of gas within the cluster environment. The above result - that galaxies have different dust-to-stellar mass ratios when separated by \hi-deficiency - shows that this physical process can affect both the atomic gas and dust properties of typical late-type galaxies. This result is not new having been shown previously by~\citet{cortese10} - they showed that the radial extent of dust was truncated for \hi-deficient galaxies. 

\subsection{Chemical Evolution}
Given that we have for a sub-sample of our galaxies measures of the mass of gas (atomic) and metals (both in dust and the gas phase) we can consider and compare the chemical evolution of cluster and filament galaxies. We have 80 filament and 18 cluster galaxies for which both HI and gas phase metallicity values are available. 

\begin{figure}
\centering
\includegraphics[width=\linewidth]{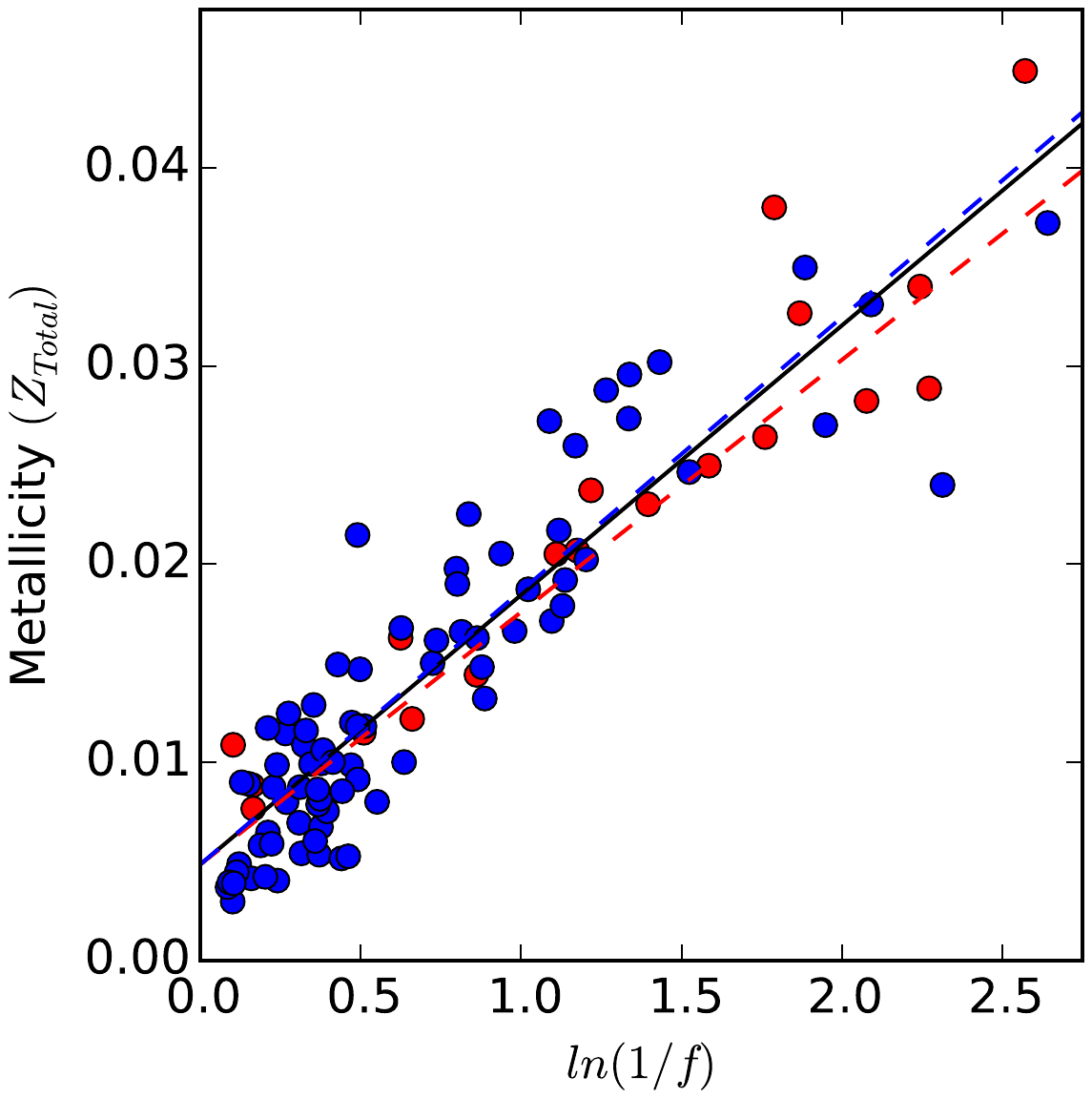}
\vspace{-2.5cm}
\caption[Chemical Evolution of Galaxies within the Coma region]{The derived metallicity (taking account of both metals in dust and the gaseous phases) against $ln(1/f)$, where $f$ is the gas fraction. Red and blue markers are used for cluster and filament galaxies, with red, blue and black lines being lines fitted to the cluster, filament and overall samples, respectively. The gradients of these lines are equal  to the effective yield $p_{eff}$.}
\label{fig:chemical_evo}
\end{figure}

\citet{Edmunds90} shows that a galaxy's metallicity $Z_{Total}=M_{metals}/M_{gas}$ is related to the gas fraction ($f = M_{gas} / (M_{stars} + M_{metals} + M_{gas})$) in the following way $Z_{Total} \le p \ln(1/f)$, where the equality applies to the closed box model i.e. one in which there are no additional inflows or outflows of gas after the galaxy has formed. $p$ is the stellar yield, which is 
the fractional mass of metals per unit mass of gas freshly formed
in nucleosynthesis. ~\citet{Edmunds90} defines the effective yield as $p_{eff} = \frac{Z_{Total}}{\ln(1/f)}$ i.e. the derived yield regardless of inflow or outflow. Essentially $p$ depends on physical processes in the interior of stars and the stellar initial mass function, while $p_{eff}$ has an additional dependence on the inflow or outflow of gas either pristine or metal enriched \citep{Davies14}. We can see if cluster galaxies have a different value of $p_{eff}$ to filament galaxies so indicating that their star forming history has been affected by the cluster environment.

In order to calculate the total gas mass of each galaxy we must take account of all the other major gas phases; molecular hydrogen, helium and the gas in the warm and hot components.~\citet{Davies14} estimate that $M^{Total}_{Gas} = 2.5 M_{HI}$ for galaxies in the Virgo cluster. This is derived from the measured HI mass, the mean ratio of molecular to atomic gas, the abundance of helium and the gas in the warm and hot components. We use this relation to adjust our \hi masses to take account of all phases of the gaseous inter-stellar medium.  

Again following \cite{Davies14} we estimate the total mass in metals to  be $M_{Metals}=1.2Z_{Gas}M_{HI}+M_{Dust}$. $Z_{Gas}$ is derived from the Oxygen abundance (12.0 + $\log_{10}$(O/H)$_{\odot}$).~\citet{Asplund01} give the solar oxygen abundance as 12.0 + $\log_{10}$(O/H)$_{\odot}$ = 8.69 and $Z_{\odot}$ = 0.014, this allows us to convert oxygen abundance (O/H) to total metallicity in the gas phase ($Z_{gas}$) using the relation $Z_{gas}$= 29.2(O/H), this assumes solar abundance ratios. However, this does not account for the metals in the warm and hot phases of the ISM.~\citet{Davies14} use the work of ~\citet{Gupta12} and estimate that the mass of metals in the hot and warm phases of the ISM to be $0.2 Z_{gas} M_{HI}$, giving $M_{Metals} = 1.2 z_{gas} M_{HI}+M_{Dust}$. The `total' metallicity ($Z_{Total}$) is then calculated from both the gaseous and dusty interstellar medium, defined as; $Z_{Total} = M_{Metals} / M^{Total}_{Gas}$. 

As an aside we find that the mean mass fraction of metals in the dust is $0.4\pm0.1$, which agrees with previous estimates of 0.5 by \cite{meyer98} and \cite{whittet92}. \cite{Davies14} performed the same analysis for galaxies in the Virgo cluster and found a value of $0.50\pm0.02$

In Figure~\ref{fig:chemical_evo} we plot the total metallicity ($Z_{Total}$) against the logarithm of the reciprocal of the gas fraction ($\ln(1/f)$). The red and blue markers and lines represent cluster and filament galaxies, respectively. We fit a linear relation with a y-axis intercept of 0, the gradient of this line is the effective yield ($p_{eff}$). The effective yield of the cluster and filament are 0.017,$\pm$\,0.01 and 0.019\,$\pm$\,0.01, respectively. Thus the two are statically identical. The black line is a fit to all the galaxies and gives an effective yield of 0.018\,$\pm$\,0.01. The galaxies appear well fitted by a linear relation. In terms of a chemical evolution model galaxies in the Coma cluster appear identical to galaxies in the filament - there is no evidence that the cluster galaxies, for example, have been stripped of gas  and so 'locking in' a fixed metalicity at a low gas fraction. 

Contrary to our inference from Figure~\ref{fig:chemical_evo} there has previously been presented some evidence that galaxies in different environments may have evolved differently. \cite{skillman1996} considered 9 galaxies in the Virgo cluster, which they then divided into 3 groups of 3 defined by their HI deficiency, which also distinguished them by their location in the cluster - the more gas rich galaxies residing in the cluster periphery. They found that these HI rich galaxies had lower metalicities than those with higher HI deficiencies and concluded that this was because they were still able to accrete metal poor gas. However, Skillman et al.'s work relied upon a very small sample size. A much larger sample has recently been considered by \cite{hughes2013}. They considered 260 galaxies from a wide range of environments from isolated to again Virgo cluster galaxies. They confirm Skillman et al.'s result that HI rich galaxies tend to have lower metalicities than those with more normal HI masses. We differ in what we have done compared to both Skillman et al. and Hughes et al. in that we do not have a measure of HI deficiency, instead we distinguish cluster and filament galaxies only by their location. We also use an estimate of the total metalicity (including dust) rather than just the $(O/H)$ gas abundance and also make an estimate of the total gas mass rather than just the atomic hydrogen. However, Hughes et al. do conclude that internal evolutionary processes are more important than environmental effects.

\section{Trends with density}
\label{sec:twd}

\begin{figure*}
\centering
\includegraphics[width=\linewidth]{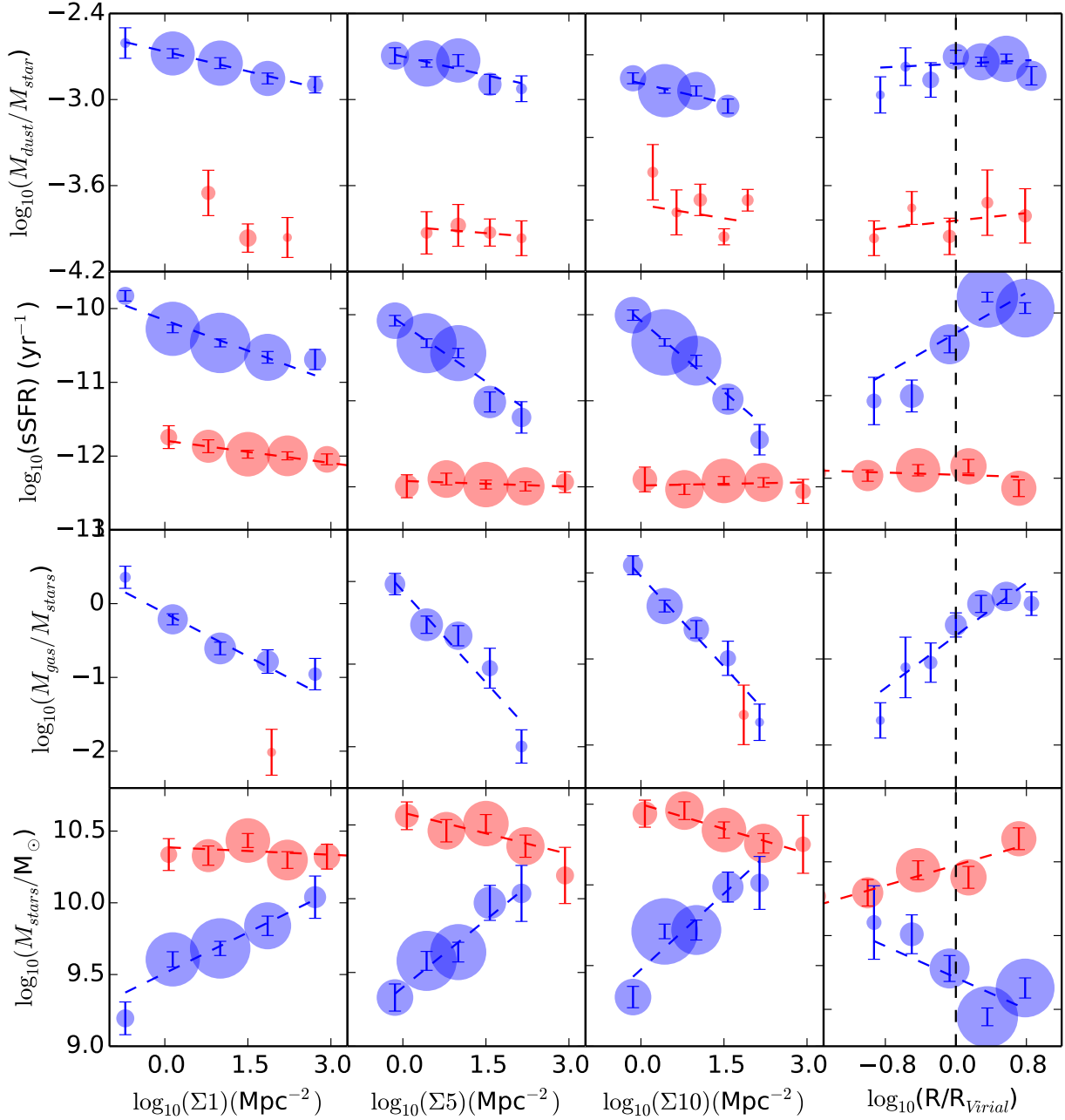}
\caption[Trends with density for all parameters]{The above shows how four different galaxy parameters change with local galaxy density as traced by $\Sigma_{1}$, $\Sigma_{5}$, $\Sigma_{10}$, and with projected cluster radius. Red and blue markers and lines are for early and late-type galaxies, respectively. Each point indicates a given density bin's mean value, and its size represents the relative number of galaxies in each bin. Each line is fitted using a $\chi^{2}$ minimisation technique.}
\label{fig:trends}
\end{figure*}

In this section we will consider how dust-to-stars\footnote{For galaxies that are detected at 250 $\mu$m, but could not be fitted with a SED because they had too few other data points, we have estimated their dust mass using the 250\mic flux density and a fixed dust temperature of 20\,K. This lets us add a further 323 galaxies to our sample, yielding a total sample of 521 galaxies with dust masses in both the filament and cluster.} and gas-to-stars ratios, sSFR\footnote{sSFR is the specific star formation rate, which is just the SFR divided by stellar mass}, stellar mass and morphological mix depend on local galaxy density ($\Sigma_{1}$, $\Sigma_{5}$, $\Sigma_{10}$, and projected cluster radius). We will also make use of the galaxy zoo morphological classifications in a slightly different way. Using the probability weightings described earlier ($p(S)$ for a spiral and $p(E)$ for an elliptical) we now define a new parameter ($\Phi$), where $\Phi = p(S) - p(E)$. This is a new non-integer way of describing a galaxy's morphology, where values of $\Phi \simeq $\,1 and -1 are expected for a definite late (spiral) and early type galaxy. As part of our trends with density analysis we will now define early type galaxies as those with negative $\Phi$ and late types as those with positive $\Phi$. Our results are summarised in Fig.~\ref{fig:trends}

For each parameter in Fig.~\ref{fig:trends} we have used evenly spaced bins over the logarithmic density range for each morphological type (early and late). Each point is the mean value in the bin with the circle size purely a relative measure of the number of galaxies in each bin. The error bar is the standard deviation in the bin divided by the square root of the number of galaxies in that bin. If a bin has less than 3 galaxies in it is discounted and neither plotted or fitted. For each morphological type, parameter vs density plot, we have fitted a straight line (e.g. $y = mx + C$), using a $\chi^{2}$ minimisation. We test whether or not each fit is consistent with a change with density (i.e. $m\ne0$). 

For each parameter and morphological type we have calculated from Fig.~\ref{fig:trends} the number of density tracers that are consistent with a change with density ($m \neq 0 $ with a significance greater than 3$\sigma$). If all 4 density tracers are consistent with a change with density then it is designated as having a `Strong' change. If 2 or 3 density tracers are consistent with a change with density then  it is designated as having a `Moderate' change. If only 1 density tracer is consistent with a change with density then it is designated as having a `Weak' change. Finally, if no density tracers are consistent with a change with density then it is designated as having  `No' change.  Clearly the different measures of density should apply over different spatial scales, but as is clear from Figure~\ref{fig:trends} trends in the data are apparent independent of the density scale used. Table~\ref{tab:summary_trends_with_density} lists each parameter and how strong (if at all) any change with density is. The bottom line is that there is pretty good evidence that late type galaxies do change their properties dependent on the local environment, but that early type galaxies do not. 

Clearly the gas-to-stars ratio of late type galaxies is strongly affected by the local density (Fig.~\ref{fig:trends}). This is consistent with the idea, discussed above, that gas is stripped from galaxies as they move through the cluster environment. Those galaxies residing in the highest density environments and/or at the smallest projected radii from the cluster centre being most affected by the gas loss. Alternately rather than gas loss the gas may have been more efficiently converted into stars. Fig.~\ref{fig:trends} also clearly shows that stellar mass is increasing with environmental density actually supporting the second alternative that gas has been taken up into stars at a faster rate in the denser cluster environments. 

\begin{table}
\centering
\caption[Strength of trends with density]{For each morphological type and parameter we measure how many density tracers are consistent with a change ($m \neq 0$) then; N=4, Strong; N=2 or 3, Moderate; N=1, Weak; and N=0 is shown in brackets.}
\begin{tabular}{ccc}
\hline
~ &Late & Early \\ \hline
Dust-to-Stars & Moderate (2) & No (0) \\
sSFR & Moderate (3) & Weak (1) \\
Gas-to-Stars & Strong (4) &  -\\
Stellar Mass & Moderate (3) & Weak (1) \\
\end{tabular}
\label{tab:summary_trends_with_density}
\end{table}

\begin{table*}
\centering
\caption[Schechter function fits to mass functions]{The tabulated values of the Schechter function fits from Figure~\ref{fig:functions}. The fits for stellar, gas (atomic) and dust mass are shown. We have caculated the mass density for each fit (see text) and listed it in the rightmost column. Refences are as follows: (1) this paper, (2)~\citet{Davies14}, (3)~\citet{Dunne11}, (4)~\citet{Panter07}, and (5)~\citet{Davies11}. }
\begin{tabular}{ccccccccc}
\hline
Sample&$\phi$&$M^{*}$&$\alpha$& $\rho$& Ref \\
~&(Mpc$^{-3}$ dex$^{-1}$)&($10^{9}$ M$_{\odot}$)&$~$& ( $10^{9}$  M$_{\odot}$ Mpc$^{-3}$) &~\\
\hline
\\
\multicolumn{6}{l}{\bf Stellar Mass \rm}\\
Coma&2.5$\pm$0.08&89$\pm$30&-1.0$\pm$0.1&220$\pm$10&(1)\\
Filament&0.05$\pm$0.01&175$\pm$48&-1.2$\pm$0.1&10.2$\pm$3.6&(1)\\
Virgo&0.3$\pm$0.1&192$\pm$117&-1.2$\pm$0.1&67$\pm$47&(2)\\
Field&0.0002$\pm$0.0001&100$\pm$20&-1.2$\pm$0.1&0.2$\pm$0.05&(4)\\
\\
\multicolumn{6}{l}{\bf Gas Mass \rm}\\
Coma&0.3$\pm$0.2&4.8$\pm$3.2&-0.7$\pm$0.4&1.3$\pm$1.2&(1)\\
Filament&0.06$\pm$0.01&3.9$\pm$1.5&-0.08$\pm$0.5&0.22$\pm$0.10&(1)\\
Virgo&0.6$\pm$0.3&4.5$\pm$1.6&-1.0$\pm$0.2&2.7$\pm$1.7&(2)\\
Field&0.009$\pm$0.001&5.0$\pm$0.1&-1.50$\pm$0.05&0.08$\pm$0.01&(5)\\
\\
\multicolumn{6}{l}{\bf Dust Mass \rm}\\
Coma&0.4$\pm$0.2&0.06$\pm$0.03&-0.8$\pm$0.2&0.02$\pm$0.02&(1)\\
Filament&0.04$\pm$0.02&0.1$\pm$0.04&-1.0$\pm$0.2&0.04$\pm$0.02&(1)\\
Virgo&0.7$\pm$0.1&0.06$\pm$0.01&-0.9$\pm$0.1&0.04$\pm$0.01&(2)\\
Field&0.006$\pm$0.001&0.040$\pm$0.004&-1.0$\pm$0.2&0.0002$\pm$0.0001&(3)\\
\hline
\end{tabular}
\label{tab:functions}
\end{table*}

\begin{figure*}
\centering
\includegraphics[width=\linewidth]{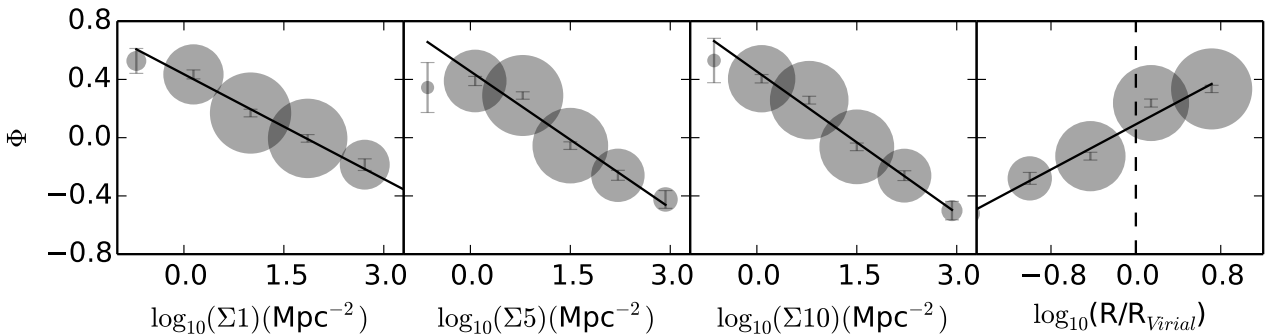}
\caption[Morphology density relation]{Morphology ($\Phi$) as a function of galaxy density as traced using $\Sigma_{1}$, $\Sigma_{5}$, $\Sigma_{10}$, and projected cluster radius (left to right).  Each marker represents a given density bin's mean value, and its size represents the number of galaxies in each bin. Each line is fitted using a $\chi^{2}$ minimisation.}
\label{fig:morphology}
\end{figure*}

We can also investigate the morphology-density relation using our new measure of morphology $\Phi$. In Figure~\ref{fig:morphology} we have calculated the change of $\Phi$ with all four environmental tracers. $\Phi$ is strongly affected by galaxy density even over the scales measured by $\Sigma_{1}$ - so if you have a near neighbour (10s to a 100 kpc away) of similar mass (we only have a small range of stellar masses in our sample, see Fig.~\ref{fig:trends}) you are more likely to be of an earlier type. Interestingly $\Phi$ is approximately equal to zero just about at the cluster Virial radius.

\section{Mass Distributions}

\begin{figure*}
\centering
\includegraphics[width=\linewidth]{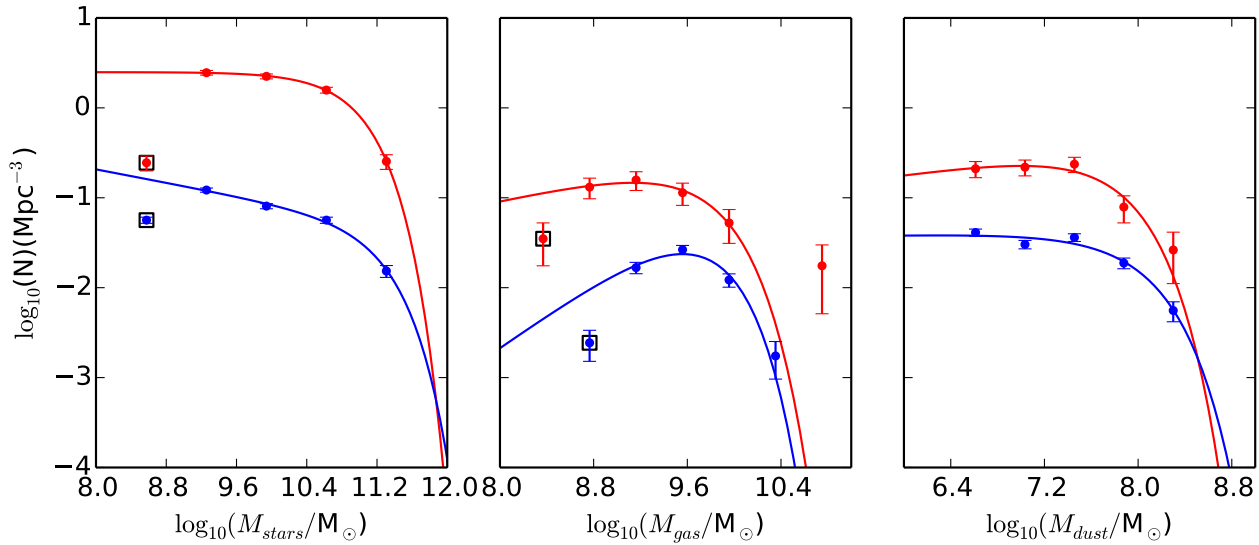}
\caption[Mass functions of the CCC]{Mass distributions for galaxies in the cluster and filament. Left-to-right are the stellar, gas (atomic) and dust mass data. The red and blue markers and lines refer to the cluster and filament samples, respectively. The lines are Schechter function fits to the data. The points with black squares were excluded from the fitting process.}
\label{fig:functions}
\end{figure*}

In order to understand how the baryonic components of galaxies change as a whole between the cluster and filament we have created stars, gas and dust mass distributions (Fig.~\ref{fig:functions}). To do this we have had to make some assumptions. Firstly, we have no information about the precise distance to individual galaxies so we assume that they are all, whether in the cluster or filament at the same distance as the cluster (105 Mpc). Secondly, ideally we would like to select by stellar mass to create the stellar mass function, atomic hydrogen mass for the gas mass function and dust mass for the dust mass function. In reality we have a complex selection of SDSS $r$ band magnitude of 17.8 for the stars, this plus a varied 21cm detection limit for the gas and the SDSS plus $Herschel$ limit for the dust - this is why we will refer to what we derive as mass distributions rather than mass functions. Thirdly, the volumes to use to get densities are rather difficult to determinate. With these provisos we will concentrate on the bright end of the mass functions, where we might hope to have completeness and on the general shape of the mass distributions rather than their difficult to determine normalisations\footnote{To over come this problem to some extent we do not fit the lowest mass point for either stellar or gas mass distributions. For our dust mass data we can be a little more confident. As stated above our dust mass limit is  $\log_{10} (M_{dust} / $M$_{\odot}) \approx 6.4$ and so we fit all data above this limit.}.

To get some rough feel for the numbers involved we estimated the volume of the cluster as a sphere with a Virial radius  of 3.1\,Mpc and hence a volume of $\sim$125 \,Mpc$^{3}$. Estimating the volume of the filament is not so straightforward because it is not a Virialised region, in the same sense as the cluster, and so does not have a well defined size. To make a crude comparison we have assumed that the filament is a sheet of depth 6\,Mpc ($\sim$2$\times R^{Coma}_{Virial}$) with a cross sectional area the same as the area covered by the NGP field at a distance of 105 Mpc (23.4 $\times 21.7$\,Mpc). We then subtract the volume of the cluster from this region yielding a volume for the filament of 2873\,Mpc$^{3}$. 

We have fitted the below Schechter function (given in terms of linear rather than logarithmic units) to each of the data sets shown in Fig.~\ref{fig:functions}: 
\begin{equation}
N (M) dM = \phi^{*} \left(\frac{M}{M^{*}} \right)^{\alpha} e^{-\frac{M}{M^{*}}} \frac{dM}{M^{*}} 
\end{equation} 
and then using: 
\begin{equation}
\rho = \phi  M^{*} \Gamma (\alpha + 2) 
\end{equation} 
calculated the mass density of each component from the derived best fitting function values, see Table~\ref{tab:functions}.

The stellar mass shown in the left panel of Figure~\ref{fig:functions} is probably the best constrained of all three mass distributions as it is derived from the complete optical sample. As noted above the best fit Schechter parameter $\phi$ is set arbitrarily by the choice of volume, but using our values the cluster is about twenty times more dense than the filament (Table~\ref{tab:functions}). The data in Table~\ref{tab:functions} also indicates that Coma is some three times more dense that Virgo and that the filament is about fifty times more dense than the general field (see references given in the caption to Table~\ref{tab:functions}). $M^{*}$ and $\alpha$ the so called `characteristic mass' and `faint end slope', are both independent of volume and thus interesting to compare. However, the values of $M^{*}$ and $\alpha$ for the cluster and filament samples are consistent ($\textless 1 \sigma$) with each other and with Virgo and the general field.  From this data there is no evidence that the environment has had any influence on the relative numbers of galaxies with different stellar masses.
  
The atomic mass distributions are shown in the central panel of Figure~\ref{fig:functions}. This is the poorest constrained of the three mass distributions as it represents the smallest fraction of the optical sample (58 and 172 galaxies in the cluster and filament, respectively). Again, with the provisos on volumes, Coma is about 6 times more dense in atomic gas than the filament - compared to the stars the gas is depleted in Coma. Most notable is the steepness of the field galaxy atomic gas mass distribution compared to Coma, filament and Virgo. The values of $M^{*}$ and $\alpha$ for the cluster and filament samples are consistent ($\textless 1 \sigma$) with each other, but the fitting errors are large because of the small numbers. 

The dust mass distribution is shown in the rightmost panel of Figure~\ref{fig:functions} and is reasonably well constrained as it represents a significant fraction of the optical sample (88 and 379 galaxies in the cluster and filament, respectively). Most notable is the large gas to dust ratio of the filament. Again, the values of $M^{*}$ and $\alpha$ for the cluster and filament samples are consistent ($\textless 1 \sigma$) with each other and with Virgo and the field. 

We can take the mass densities ($\rho$) from Table~\ref{tab:functions} and create ratios that are usefully independent of the volume used. We have created one for each of the four environments (Coma, filament, Virgo, and the field), see Table~\ref{tab:ratios}.

\begin{table}
\centering
\caption[Mass ratios for various environments]{The ratios of mass density (from Table~\ref{tab:functions}) within each environment. We have normalised each ratio to stellar mass density.}
\begin{tabular}{ccc}
\hline
Sample & $\log_{10}(\rho_{gas}/\rho_{stars})$  & $\log_{10}(\rho_{dust}/\rho_{stars})$ \\
\hline
Coma&-2.2\,$^{+0.3}_{-1.3}$&-4.0\,$^{+0.2}_{-0.5}$\\
Filament&-1.7\,$^{+0.2}_{-0.4}$&-3.4\,$^{+0.2}_{-0.6}$\\
Virgo&-1.4$\,^{+0.3}_{-1.2}$&-3.2\,$^{+0.2}_{-0.6}$\\
Field&-0.5\,$^{+0.1}_{-0.1}$&-3.0\,$^{+0.1}_{-0.2}$\\\hline
\end{tabular}
\label{tab:ratios}
\end{table}

The Coma, filament and Virgo environments are all statically the same (Table~\ref{tab:ratios}) in terms of $\log_{10}(\rho_{gas}/\rho_{stars})$ and $\log_{10}(\rho_{dust}/\rho_{stars})$, with the exception of $\log_{10}(\rho_{gas}/\rho_{stars})$ in the field - the field is much more gas rich than any of the other three environments. However, all four environments are statistically the same with regard to $\log_{10}(\rho_{dust}/\rho_{stars})$. These results again imply that gas is affected by higher density environments much more so than the other significant component of the interstellar medium, dust. We have ordered the mass density ratios in Table~\ref{tab:ratios} from lowest to highest and they correspond to what we might have expected with the possible exception of the ordering of the filament and Virgo environments.

Given that we have calculated the baryonic mass densities of the Coma cluster (in galaxies) we can also calculate the total mass of each component  using our above estimate of its volume (125\,Mpc$^{3}$). As we have also measured the velocity dispersion of the cluster, we can measure the Virial mass of the cluster using:
\begin{equation}
M_{Virial} = \frac{ 5 R_{virial} \sigma^{2} } {G} 
\end{equation} 
Where $M_{Virial}$, $R_{Virial}$, $\sigma$, and $G$ are the Virial mass, Virial radius, velocity dispersion and gravitational constant, respectively.  In Table~\ref{tab:cluster_mass} we compare the measured baryonic mass of material in galaxies with the total dynamical mass of the cluster. The overall ratio of dynamical mass to baryonic mass (excluding the x-ray gas) of the cluster is $\sim104$. 

\begin{table}
\centering
\caption[Mass parameters for the Coma Cluster]{Mass parameters for the Coma Cluster derived from optically selected galaxies}
    \begin{tabular}{cc}
\hline
    Component  & Mass (M$_{\odot}$) \\
\hline
    Total Stellar Mass & 2.8$\times 10^{13}$  \\
    Total Gas Mass & 1.6$\times 10^{11}$  \\
    Total Dust Mass & 2.5$\times 10^{9}$  \\ \hline
    Total Baryonic (Excluding X-ray gas)  & 2.8$\times 10^{13}$  \\
    Virial Mass & 2.9$\times 10^{15}$  \\ 
\hline    
\end{tabular}
\label{tab:cluster_mass}
\end{table}

\section{Summary}
We have undertaken a $Herschel$ far infrared survey of the Coma cluster and the galaxy filament it resides within.  Our survey covers an area of $\sim$150 deg$^2$ observed in five bands  at 100, 160, 250, 350 and 500\,$\mu$m. We have used the SDSS spectroscopic survey to define an area and redshift selected sample of 744 Coma cluster galaxies - the Coma Cluster Catalogue (CCC). For comparison we also define in a similar way a sample of 951 galaxies in the connecting filament - the Coma Filament Catalogue (CFC).

We have used the optical positions and parameters of these CCC and CFC galaxies to define appropriate apertures to measure their far infrared emission. We have detected 99 of 744 (13\,\%)  and 422 of 951 (44\,\%) of the cluster and filament galaxies in the SPIRE 250\,$\mu$m band, which we use for our initial far infrared selection. We have carried out simulations to ensure that our process of associating far infrared and optical sources leads to less than 5\% mis-identified objects.

In order to better understand the global detection rates we have separated the cluster and filament galaxies into 3 morphological categories using the Galaxy Zoo data: early, $p(E) \textgreater 0.8$; late, $p(S) \textgreater 0.8$ and uncertain where $p(E) \textless 0.8$ and $p(S) \textless 0.8$. We examined the detection rate for each morphological group in the 250\,$\mu$m band as it has the highest detection rate of all the $Herschel$ bands. Early and late-type galaxies have the same relative detection rates in the cluster and filament, but overall the fraction of galaxies detected in the cluster is lower because of the morphological mix. The relative detection rates of our 'uncertain' types (S0/Sa?) is noticeably different between the filament (higher rate) and the cluster (lower rate). This is one of the few clear measurements of a difference between the filament and cluster environment.

We compare the cluster velocity dispersions of detected and undetected late, uncertain and early-type galaxies. We found that all morphological types have velocity dispersions statically identical to each other and the overall cluster value. This implies that all morphological types have been in the cluster longer than a crossing time and hence longer than the timescale for ram pressure stripping - the two of which are comparable. 

For galaxies detected in all five far infrared bands we fit a modified blackbody with a fixed emissivity index of $\beta=2$, giving dust masses and temperatures for 198 galaxies. As expected early-type galaxies show lower dust masses than late-type galaxies, but they tend to have hotter dust temperatures. When comparing the far infrared properties of galaxies in Coma and the filament we find that they are statistically identical with the exception of the hotter temperatures of early type cluster galaxies.

Combining our data with that from the literature we estimate the total gas mass as well as the mass in metals (in the gas phase and in the dust) for a sub-sample of Coma and filament galaxies. We find no evidence for a difference in the effective yields of these two populations. This implies that neither population has been subjected to gas loss or infall to a greater or lesser extent than the other population.  

We have compared a number of galaxy parameters (dust-to-stars, gas-to-stars, sSFR, and morphology) for Coma cluster and filament galaxies  and considered how each is affected by local galaxy number density ($\Sigma_{1}$, $\Sigma_{5}$, $\Sigma_{10}$, and projected cluster radius). Using the galaxy zoo weightings of probability of being a spiral ($p(S)$) or elliptical ($p(E)$) we defined a new parameter ($\Phi$), where $\Phi = p(S) - p(E)$.  We show that $\Phi$ is strongly affected by local number density (environment), which is another way of representing the morphology-density relation~\citep{dressler80}. We find that the above measured parameters of early-type galaxies appear to be only very weakly or not at all affected by their environment. Late types also show a weak change in most of the above parameters with environment - the exception being gas-to-stars ratio, which is strongly affected by environmental density on all scales. 

In order to understand how the baryonic components of the galaxies change as a whole between the cluster and filament we have created baryonic mass functions. Using Schechter fits to stellar, gas and dust mass functions for the Coma cluster and filament we compare their parameters ($\phi$, $M^{*}$ and $\alpha$) with values found for the Virgo cluster and a field sample. Other than $\phi$, which is dependant on a somewhat arbitrary calculation of volume, we found $M^{*}$ and $\alpha$ to be identical between environments. We calculate the ratio of gas-to-stellar and dust-to-stellar mass densities for each environment finding that within the errors Virgo, Fornax, Coma and the filament are all gas deficient when compared to the general field, but that this is not so for the dust. Galaxies in dense environment are far more prone to gas rather than dust loss.

\vspace{0.5cm}
\begin{center}
{\bf Acknowledgements} \\
\end{center}

The  H-ATLAS is a project with $Herschel$, which is an
ESA space observatory with science instruments provided
by European-led Principal Investigator consortia and with
important participation from NASA. The H-ATLAS web-
site is http://www.h-atlas.org/. 

LD, SM and RI acknowledge support from the ERC in the form of the Advanced Investigator Program, COSMICISM. LD and SM acknowledge support from the European Research Council in the form of Consolidator Grant CosmicDust. 

Funding for the Sloan Digital Sky Survey IV has been provided by
the Alfred P. Sloan Foundation, the U.S. Department of Energy Office of
Science, and the Participating Institutions. SDSS-IV acknowledges
support and resources from the Center for High-Performance Computing at
the University of Utah. The SDSS web site is www.sdss.org.

SDSS-IV is managed by the Astrophysical Research Consortium for the 
Participating Institutions of the SDSS Collaboration including the 
Brazilian Participation Group, the Carnegie Institution for Science, 
Carnegie Mellon University, the Chilean Participation Group, the French Participation Group, Harvard-Smithsonian Center for Astrophysics, 
Instituto de Astrof\'isica de Canarias, The Johns Hopkins University, 
Kavli Institute for the Physics and Mathematics of the Universe (IPMU) / 
University of Tokyo, Lawrence Berkeley National Laboratory, 
Leibniz Institut f\"ur Astrophysik Potsdam (AIP),  
Max-Planck-Institut f\"ur Astronomie (MPIA Heidelberg), 
Max-Planck-Institut f\"ur Astrophysik (MPA Garching), 
Max-Planck-Institut f\"ur Extraterrestrische Physik (MPE), 
National Astronomical Observatory of China, New Mexico State University, 
New York University, University of Notre Dame, 
Observat\'ario Nacional / MCTI, The Ohio State University, 
Pennsylvania State University, Shanghai Astronomical Observatory, 
United Kingdom Participation Group,
Universidad Nacional Aut\'onoma de M\'exico, University of Arizona, 
University of Colorado Boulder, University of Oxford, University of Portsmouth, 
University of Utah, University of Virginia, University of Washington, University of Wisconsin, 
Vanderbilt University, and Yale University.

The $Herschel$ spacecraft was designed, built, tested, and launched under a contract to ESA managed by the Herschel/Planck Project team by an industrial consortium under the overall responsibility of the prime contractor Thales Alenia Space (Cannes), and including Astrium (Friedrichshafen) responsible for the payload module and for system testing at spacecraft level, Thales Alenia Space (Turin) responsible for the service module, and Astrium (Toulouse) responsible for the telescope, with in excess of a hundred subcontractors.

This research has made use of the GOLDMINE Database.

E. Ibar acknowledges funding from CONICYT/FONDECYT postdoctoral project N$^\circ$:3130504.
 
\bibliographystyle{mn2e} 
\bibliography{references}

\end{document}